\documentclass[11pt,a4paper]{article}
\usepackage[utf8]{inputenc}
\usepackage{amsmath}
\usepackage{amsthm}
\numberwithin{equation}{section}
\usepackage{amsfonts}
\usepackage{mathtools}
\usepackage{mathrsfs}
\usepackage{mathpazo}
\usepackage{multirow}
\usepackage{amssymb}
\usepackage{graphicx}
\usepackage{ifpdf}
\usepackage{braket}
\usepackage{cancel}
\usepackage[dvipsnames]{xcolor}
\usepackage{multicol}

\usepackage{cite}
\usepackage[bookmarks=true,colorlinks=true,linkcolor=black,citecolor=orange,urlcolor=orange,bookmarksnumbered]{hyperref}

\usepackage[left=2.5cm, right=2.5cm, top=2.5cm, bottom=3cm]{geometry}

\usepackage[normalem]{ulem}

\usepackage{accents}
\newcommand{\ubar}[1]{\underaccent{\bar}{#1}}

\newcommand{\bI}{\bar{I}}
\newcommand{\bJ}{\bar{J}}
\newcommand{\bK}{\bar{K}}
\newcommand{\bL}{\bar{L}}
\newcommand{\bM}{\bar{M}}
\newcommand{\bN}{\bar{N}}
\newcommand{\bP}{\bar{P}}

\newcommand{\ubI}{\ubar{I}}
\newcommand{\ubJ}{\ubar{J}}
\newcommand{\ubK}{\ubar{K}}
\newcommand{\ubL}{\ubar{L}}
\newcommand{\ubM}{\ubar{M}}
\newcommand{\ubN}{\ubar{N}}
\newcommand{\ubP}{\ubar{P}}

\newcommand{\diff}{\mathrm{d}}

\begin{document}

\renewcommand*{\thefootnote}{\fnsymbol{footnote}}
\begin{center}

{\LARGE \bf Current Algebra and Generalised Cartan Geometry}

\vspace{0.5truecm}

{Falk Ha\ss ler\textsuperscript{1)}, Ondřej Hulík\textsuperscript{2)} and David Osten\textsuperscript{1)}}

\vspace{0.5truecm}

{\textsuperscript{1)}\em Institute for Theoretical Physics (IFT), University of Wroc\l aw \\
pl. Maxa Borna 9, 50-204 Wroc\l aw, Poland
}

\vspace{0.2truecm}

{{\tt \{falk.hassler,david.osten\}@uwr.edu.pl}}

\vspace{0.5truecm}

{\textsuperscript{2)}\em Institute for Mathematics 
 Ruprecht-Karls-Universitat Heidelberg,
 \\
 69120 Heidelberg, Germany 
}

\vspace{0.2truecm}

{\tt ondra.hulik@gmail.com}

\vspace{0.5truecm}
\end{center}
\begin{abstract}
This article shows that the approach to generalised curvature and torsion pioneered by Poláček and Siegel \cite{Polacek:2013nla} is a generalisation of Cartan Geometry -- rendering latter natural from the point of view of O$(d,d)$-generalised geometry. We present this approach in the generalised metric formalism and show that almost all parts of the additional higher generalised tensors appearing in this approach correspond to covariant derivatives of the generalised Riemann tensor. As an application, we use this framework to phrase $\sigma$-model dynamics in an explicitly covariant way -- both under generalised diffeomorphisms and local gauge transformations.
\end{abstract}

\vspace*{-0.5cm}

\renewcommand*{\thefootnote}{\arabic{footnote}}
\setcounter{footnote}{0}

\tableofcontents


\section{Introduction} 
Mathematics has often successfully served as guiding principle for physics. Just think about the advent of general relativity, which revealed an intriguing relation between gravitational interaction and the pseudo-Riemannian geometry of space and time. However, recently it became more common that a thorough analysis of physical systems may lead to new idea in mathematics. 

The physical system of interest here, non-linear $\sigma$-models \cite{GellMann:1960np}, have a proven track record as a mediator between standard methods in physics and geometry. One important example being the relation between the Ricci-flow of the target space and the RG-flow of the non-linear $\sigma$-model \cite{Friedan:1980jf}. One of the latter captures their dynamics in terms of a flow on their phase space generated by a function called the Hamiltonian. This only works, because the phase space is equipped with a symplectic structure encoded in terms of Poisson brackets. While this formalism is the canonical starting point for a quantisation and the analysis of conserved quantities, it comes with the drawback that the target space geometry is more opaque in comparison to the Lagrangian approach.

Non-linear $\sigma$-models are theories of fields that are embeddings of one manifold, here called world volume, into another manifold $M$, called target space, in following. This article is motivated by the question
\begin{center}
    ``How is the geometry of the target space reflected in the phase space and the Hamiltonian of two-dimensional (bosonic) $\sigma$-models?''
\end{center}
Remarkably, from this questions one is lead to a generalisation of Riemannian geometry dubbed generalised geometry. Besides its rich mathematical structure \cite{Hitchin:2003cxu,Gualtieri:2003dx}, it also has proven very useful in exploring physical phenomenal like integrability \cite{Lust:2018jsx,Demulder:2018lmj,Borsato:2021gma,Borsato:2021vfy} and dualities \cite{Severa:2015hta,Hassler:2017yza,Sakatani:2019jgu,Osten:2019ayq}. However, despite its success, generalised geometry is not yet completely en par with its ancestor. A long standing problem is to find a generalised version of the Riemann tensor. There are proposals \cite{Jeon:2010rw,Coimbra:2011nw,Hohm:2011si} but they are plagued with undetermined components. This shortcoming is still accepted because the undetermined components drop out from physical observables, at least in the classical regime. Even more troublesome is that the generalised curvature is not manifest on the phase space and, consequently, in its dynamics. 
Viewing these problems in the light of lessons learned from the study of dualities leads us to the conclusion that there should be an extension of generalised geometry which naturally describes the phase space structure and at the same time cures at least some of the aforementioned problems.

The extension in question originates from Cartan\footnote{Ideas leading to the current understanding of Cartan geometry go back more than a century to Cartan's re-formulation of Riemannian geometry in terms of co-frames \cite{Cartan1922}. As a first introduction to the subject, we found the lecture notes \cite{Alexandre2024} 
very helpful.}, which on its own is an efficient way of capturing curvatures. Therefore, we call it \textit{generalised Cartan geometry}. As this is a new framework, we will first define it by lifting definitions of standard Cartan geometry to generalised geometry and then show that it is implemented naturally on the phase space of two-dimensional, non-linear $\sigma$-models. Our main starting point is work by Poláček and Siegel \cite{Polacek:2013nla,Polacek:2017hnq}. They used intuition from standard Cartan geometry for a natural derivation of curvature and torsion that are covariant under the generalised diffeomorphisms of O$(d,d)$ generalised geometry. This involves an additional extension of the Cartan model space and new auxiliary parts of the connection, and also yields additional generalised tensors apart from generalised Riemann tensor and the generalised torsion. In hindsight, one might understand their approach as index-version\footnote{For the physicist authors, this language is more familiar and hence dominates the rest of the article.} of symplectic reduction of Courant algebroids \cite{Bursztyn:2005vwa,Severa:2017oew} that have been discussed by mathematicians before. Remarkably, these reductions can be used to explain dualities which relate different target spaces geometries by establishing a canonical transformation between the corresponding $\sigma$-model phase spaces \cite{Severa:2015hta,Severa:2018pag,Butter:2022iza}. We argue that all these approaches secretly implement a generalisation of Cartan geometry that is natural from the point of view of generalised geometry. The corresponding objects are summarised in table \ref{tab:Dictionary}.

To make a clear connection to already existing results in the literature, we discuss them in the generalised metric formalism. For metric compatible and generalised torsion-free connections, it turns out that most projections of the new tensors on the right side of the table's last column actually correspond to projections of the generalised Riemann tensor or its covariant derivative. Similarly to the generalised Riemann tensor, these higher tensors depend both on the physical and undetermined, unphysical part of the connection. However, there are projections that constitute a new covariant tensor, in the sense that is not expressible in terms of generalised Riemann tensor.
    
This framework allows to phrase string world-sheet dynamics in an explicitly covariant way -- both under generalised diffeomorphisms and local gauge (double Lorentz) transformations. Thus, we conclude that generalised Cartan geometry provides a new link between the phase space structure and the geometry of non-linear $\sigma$-models. It generalises recent work by Lacroix \cite{Lacroix:2023ybi} for generalised cosets to arbitrary backgrounds and gauge symmetries. As expected, one notices that the string dynamics depends neither on the new auxiliary parts of the generalised Cartan connection, nor its part undetermined by generalised metric compatibility.

The article is organised into three major parts. First, we review the most important definitions of Cartan geometry and show how they can be realised in the phase space of a point particle. After a short introduction to generalised geometry, we justify our lift of these definitions to the generalised tangent space. As the main guiding principle, we use here insights from generalised dualities to extend the construction of Poláček and Siegel. Finally we show that the newly discovered structures find a very simple realisation on the $\sigma$-model phase space in terms of a current algebra. The fundamental dynamic object is this first part is a frame. However, generalised geometry was originally formulated in terms of a generalised metric. Thus, to analyse the now objects we found further, we transition in section~\ref{chap:genMetricFormalism} to a generalised metric formulation. In particular, we analysis which of the new quantities we revealed are related to already known ones and which of them are genuinely new. Coming back to physics, we show how generalised Cartan geometry permits to formulate the dynamics of strings in a fully duality covariant way in section~\ref{chap:CovariantStringDynamics}. While section~\ref{chap:Outlook} concludes the paper with an outlook on possible direction of future exploration.

\begin{table*}
\centering \small
    \begin{tabular}{c|cc}
        		& Cartan Geometry & Generalised Cartan Geometry \\ \hline
      	phase space realisation & point particle on $M$ & bosonic string on $M$\\
       	gauge symmetry & \multicolumn{2}{c}{Lie group $H$, Lie algebra $\mathfrak{h}$} \\
       	underlying bundle & \multicolumn{2}{c}{principle $H$-bundle $\pi: \ P \rightarrow M$} \\
        model algebra & Lie algebra $\mathfrak{g} \supset \mathfrak{h}$ & quasi-Lie bialgebra $\mathfrak{d}\supset \mathfrak{h}$ \\
        underlying algebroid bracket & $\mathfrak{g}$-twisted Lie bracket $[ \ , \ ]_{L,\mathfrak{g}}$ & $\mathfrak{d}$-twisted Dorfman bracket $[ \ , \ ]_{D,\mathfrak{d}}$ \\ \hline
        \textbf{Cartan connection} & isomorphism $\theta: \ TP \rightarrow \mathfrak{g}$ & $\eta$-isomorphism $\theta: \ (T \oplus T^\star)P \rightarrow \mathfrak{d}$ \\ \hline
        independent components & frame ${e_a}^m$ & generalised frame ${E_A}^M$ \\
        		& spin connection ${\omega_m}^\alpha$ & generalised spin connection ${\Omega_M}^\alpha$ \\
        		& & Poláček-Siegel field $\rho^{\alpha \beta}$ \\ \hline
        \textbf{Cartan curvature} & $\Theta_{\mathcal{A}\mathcal{B}} = - [\theta_{\mathcal{A}} , \theta_{\mathcal{B}}]_{L,\mathfrak{g}} \in \mathfrak{g}$ & $\Theta_{\mathcal{A}\mathcal{B}} = - [\theta_{\mathcal{A}} , \theta_{\mathcal{B}}]_{D,\mathfrak{d}} \in \mathfrak{d}$ \\ \hline
        independent components & torsion ${T^m}_{kl} $ & generalised torsion $\mathcal{T}_{KLM}$ \\
        & Riemann tensor ${R_{kl,m}}^n$ & generalised Riemann tensor ${\mathcal{R}_{KL,MN}}$ \\
        & & '$\rho$-torsion' $\mathcal{A}_{KL,MN}$ \\
        & & 'higher curvature' $\mathcal{F}_{P,KL,MN}$
    \end{tabular}
    \label{tab:Dictionary}
    \caption{Corresponding objects of standard and generalised Cartan Geometry over a manifold $M$. Let us note that, in view of its generalisation, we work with a trivially extended version of Cartan curvature on the full $TP$, without projection to $TM$. This is introduced in detail in section \ref{chap:CartanGeometryOrdinary}.}
\end{table*}

\section{Cartan geometry} \label{chap:CartanGeometryOrdinary}
Before discussion generalisations of Cartan geometry, we will review its standard form.
\subsection{Cartan connection and curvature}
Consider a $d$-dimensional smooth manifold $M$. Cartan geometry offers a very general setup to describe connections and their curvatures over $M$. Roughly speaking, Cartan geometry models the tangent space at $T_x M$ at each point $x$ of $M$ as the quotient of two Lie algebras $\mathfrak{g}/\mathfrak{h}$. Each element of this coset describes a translation along the tangent space, while the group $H$ (associated to $\mathfrak{h}$) encodes rotations. As the latter can be different in different points of the $T_x M$, $H$ will play the role of a gauge symmetry in the physical set-up. As central and canonical example in context of this article serves a metric-compatible affine connection with or without torsion. In this case the group $G$ associated to $\mathfrak{g}$ is the Euclidean group, $H$ implements rotations and the coset $G/H \sim \mathbb{R}^{d}$ described a flat plane. In the same vein, Minkowski space arises from a quotient of the Poincar\'e group by Lorentz transformations.

For the purposes of this paper, it is sufficient to consider local properties based on a local chart $U \subset M$ with coordinates $x^\mu$. Typical accounts of Cartan geometry have also requirements about global properties, see \cite{cap2009parabolic}. In order to give meaning to generalisation that are performed in section \ref{chap:GeneralisedCartanGeometry}, we first give a working definition of Cartan geometry.
\begin{enumerate}
    \item The underlying objects are:
    \begin{itemize}
        \item A Lie group $H$, called \textit{gauge group}, with Lie algebra $\mathfrak{h}$ and its associated principal bundle
        \begin{equation}
            \pi: P \rightarrow M, \quad \text{locally: } P \vert_U = H \times U. \label{eq:PrincipalBundleDefinition}
        \end{equation}
        The prototypical example here will be that $H$ is O$(d)$ or the Lorentz group O$(1,d-1)$. 
        
        \item A $d$-dimensional (coset) model space $\frac{G}{H}$, formed by a (reductive) quotient of the Lie group $G$ and $H$. For the presentation here, only its Lie algebra $\mathfrak{g}$, which we will dub \textit{model algebra} in the following, is relevant. In the prototypical example where $H$ is the group of rotations or the Lorentz group, the model algebra $\mathfrak{g}$ would be the Euclidean or Poincar\'e algebra, where $\mathfrak{g} = \mathbb{R}^d \rtimes  \mathfrak{o}(d)$ (or $\mathfrak{g} = \mathbb{R}^{d} \rtimes  \mathfrak{o}(1,d-1)$).
    \end{itemize}
    \item A \textit{Cartan connection} $\theta$ is a fibrewise isomorphism of $TP$ and the model algebra $\mathfrak{g}$
    \begin{equation}
        \theta(p): \ T_p P \rightarrow \mathfrak{g}. \label{eq:CartanConnectionDefinition}
    \end{equation}
    In order for this to describe an identification $T_{\pi(p)} M \simeq \mathfrak{g}/\mathfrak{h}$, we assume the following additional properties for $\theta$:
   \begin{itemize}
        \item Left-invariant vector fields of $TH$ are identified with the subalgebra $\mathfrak{h} \subset \mathfrak{g}$, via absolute parallelism. Assuming that we denote them by $X_\xi$, where $\xi\in\mathfrak{h}$, the Cartan connection has to satisfy
        \begin{equation}\label{eq:NormalCartanIdentity}
            \theta(X_\xi) = \xi\,.
        \end{equation}
        \item \textit{Equivariance}. The right-action of $h \in H$, which we denote by $R_h$ here and later refer to as $H$-gauge transformation, has to match the adjoint action on $\mathfrak{h}$ through
        \begin{equation}
            R^*_h \theta = \mathrm{Ad}_{h^{-1}} \theta\,. \label{eq:Equivariance}
        \end{equation}
        This condition is result of the fact that the identification $T_{\pi(p)} M \simeq \mathfrak{g}/\mathfrak{h}$ is not $H$-invariant. We will interpret this as gauge transformations for the components of $\theta$. 
    \end{itemize}
    Targeted at its generalisation later and connecting to the conventions in previous literature \cite{Polacek:2013nla,Butter:2022iza}, let us give we give an explicit expression for $\theta$. Let $\mathcal{A}=(\alpha,a),\mathcal{B}=(\beta,b),...$ denote indices on $TP = TH \oplus T_x U \simeq \mathfrak{h} \oplus T_x U$ for a in general non-holonomic basis $(\nabla_\alpha , \partial_a = {e_a}^m \partial_m)$ of $TP$, in particular such that $X_\xi = \xi^\alpha \nabla_\alpha$ corresponds to right-invariant vector fields on $H$ used in the definition above. With these choices a general expression for $\theta$ in accordance with the above definition of $\theta$ is:
    \begin{equation}
    {\theta_{\mathcal{A}}}^{\mathcal{M}}(x) = \left( \begin{array}{cc}
        \delta_\alpha^\mu & 0  \\
        {\omega_a}^{\mu}(x) & {e_a}^m(x) 
    \end{array} \right)\,. \label{eq:CartanConnectionChoice}
    \end{equation}
Where $\mathcal{K}=(\kappa,k),\mathcal{L}=(\lambda,l)$ are indices on $\mathfrak{g} = \mathfrak{h} \oplus \mathfrak{g}/\mathfrak{h}$.

     The non-trivial part of the Cartan connection decomposes into a connection $\omega_a \in \mathfrak{h}$ and the frame field ${e_a}^m$ on $M$. An important consequence of equivariance of $\theta$ is that these non-trivial components need to transform under $H$-gauge transformations. The explicit expression are discussed below in \eqref{eq:OrdHTensors}.
     
     In comparison to some other accounts of Cartan geometry, \eqref{eq:CartanConnectionChoice} could be labeled as an 'extended' Cartan connection, due to fact that we include the components ${\theta_{\alpha}}^\mathcal{M}$. 
    
    \item The \textit{Cartan curvature} $\Theta$ for a Cartan connection $\theta: \ TP \rightarrow \mathfrak{g}$ is:
        \begin{equation}
            \Theta = - \mathrm{d} \theta + \frac{1}{2} [\theta, \theta] \in \mathfrak{g} \label{eq:CartanCurvatureNormal}
        \end{equation}
        where $\theta$ is understood as $\mathfrak{g}$-valued 1-form on $P$. The non-trivial components \linebreak $\Theta^{\mathcal{M}}{}_{ab} = \left({R^{\mu}}_{ab} , T^{m}{}_{ab} \right)$ are the standard expressions for torsion and curvature
        \begin{align}
            T &= - \mathrm{d} e + [\omega, e ] \in \mathfrak{g}/\mathfrak{h} \label{eq:TorsionStandard} \\
            R &= - \mathrm{d}\omega + \frac{1}{2}[\omega , \omega] \in \mathfrak{h} \label{eq:CurvatureStandard}
        \end{align}
        after being pulled back to the physical space $M$. When $\mathfrak{g}$ is the Poincar\'e or Euclidean algebra, this produces the standard expressions for torsion $T$ and Riemann curvature $R$ of an affine connection in the frame formulation. In our non-holonomic basis, the remaining non-vanishing components, $\Theta_{\alpha \beta}{}^\gamma$ and $\Theta_{\alpha b }{}^c$, in this slightly extended version of Cartan geometry correspond to structure constants of $\mathfrak{h}$ and the action of $\mathfrak{h}$ on the $T_{\pi(p)}M$
\end{enumerate}
\paragraph{Derivation from an extended space.}
Our definitions here have been motivated to match the later analysis of point particle Poisson algebra in the next section. There, a non-holonomic basis of $T H$ and covariant transformation rules for tensors appear naturally and have the advantage that all relevant quantities are independent of the fibre's coordinates. However, one can also take an alternative route relying on the standard coordinate basis for $T H$. In this case, we consider the group elements $h \in H$ whose adjoint action on $\mathfrak{g}$ is denoted by
\begin{equation}\label{eq:DefTildeMg}
  \mathrm{Ad}_{h^{-1}} t_{\mathcal{L}} = h^{-1} t_{\mathcal{L}} h := \widetilde{M}_{\mathfrak{g}}(t _\mathcal{L}) = (\widetilde{M}_{\mathfrak{g}})_{\mathcal{L}}{}^{\mathcal{K}} t_{\mathcal{K}}\,.
\end{equation}
$\widetilde{M}_{\mathfrak{g}}$'s definition is chosen on purpose such that
\begin{equation}
  R_{h'} \widetilde{M}_{\mathfrak{g}} = Ad_{h'^{-1}} \widetilde{M}_{\mathfrak{g}}
\end{equation}
holds which mimics the equivariance condition \eqref{eq:Equivariance}. Moreover, we will need the right-invariant Maurer-Cartan form $\widetilde v = \mathrm{d} h h^{-1} $, i.e. $R^\star_{h'} \widetilde{v} = \widetilde{v}$.
Finally, we assume that the vector fields $X_\xi$ introduced above act as
\begin{equation}
  \iota_{X_\xi} \widetilde{v} = \mathrm{Ad}_h \xi\,.
\end{equation}
This relation implies that $X_\xi$ are the dual vector fields of the left-invariant Maurer-Cartan form
\begin{equation}
  \widetilde{e} = h^{-1} \mathrm{d} h
  \qquad \text{with} \qquad
  \widetilde{v} = \mathrm{Ad}_h \widetilde{e}\,,
  \qquad \text{and} \qquad
  \iota_{X_\xi} \widetilde{e} = \xi\,.
\end{equation}
Using Cartan's magic formula, one gets $\iota_{[ X_{\xi_1}, X_{\xi_2} ]} \widetilde{e} = \iota_{X_{\xi_1}} \iota_{X_{\xi_2}} \mathrm{d} \widetilde{e} = \iota_{X_{[\xi_1, \xi_2]} } \widetilde{e}$ and therefore verify their defining property $[ X_{\xi_1}, X_{\xi_2}] = X_{[\xi_1, \xi_2]}$. As $\widetilde v$ is only valued in $\mathfrak{h}$, we embed it by
\begin{equation}\label{eq:DefTildeVg}
  \widetilde{V}_{\mathfrak{g}} = \widetilde{v} \oplus \mathbf{1}_{\mathfrak{g}/\mathfrak{h}}
\end{equation}
as the action on the full Lie algebra $\mathfrak{g}$. Equipped with $\widetilde{M}_{\mathfrak{g}}$ and $\widetilde{V}_{\mathfrak{g}}$, we eventually define the Cartan connection as
\begin{equation}\label{eq:twistForCartan}
  \widehat{\theta} = \widetilde{M}_{\mathfrak{g}} \overline{\theta} \widetilde{V}_{\mathfrak{g}}
\end{equation}
where $\overline{\theta}$ has exactly the same form as indicated in \eqref{eq:CartanConnectionChoice}. Again, we require that it may only depend on the coordinates of the base manifold and conclude that it will not change under any fibre transformations like $R_h \overline{\theta} = \overline{\theta}$. Combining this property with the ones established above, we obtain
\begin{equation}\label{eq:EquivarianceExplicit}
  R^*_h \widehat{\theta} = R_h \widetilde{M}_{\mathfrak{g}} R_h \overline{\theta} R^*_h \widetilde{V}_{\mathfrak{g}} = \mathrm{Ad}_{h^{-1}} \widetilde M_{\mathfrak{g}} \overline{\theta} \widetilde{V}_{\mathfrak{g}} = \mathrm{Ad}_{h^{-1}} \widehat{\theta}
\end{equation}
and thereby prove equivariance \eqref{eq:Equivariance}. In the same vein, we establish \eqref{eq:NormalCartanIdentity}, namely
\begin{equation}\label{eq:CartanIdentityExplicit}
  \iota_{X_\xi} \widehat{\theta} = \mathrm{Ad}_{h^{-1}} \overline{\theta} \mathrm{Ad}_{h} \xi = \xi\,,
\end{equation}
At this point, we seen that all properties of the Cartan connections arise naturally from the twists performed by $\widetilde{M}_{\mathfrak{g}}$ and $\widetilde{V}_{\mathfrak{g}}$. Both only depend on the subalgebra $\mathfrak{h}$ and the choice of complement $\mathfrak{g}/\mathfrak{h}$ inside $\mathfrak{g}$.


\subsection{Realisation of Cartan geometry on the point particle phase space}\label{chap:PointParticle}
The structure of Cartan geometry is naturally realised on the phase space of a point particle. Let $p_m$ and $x^m$ be the canonical momentum and coordinate field of a particle moving in $M$, with the conventions $\{p_m , x^n \} = - \delta_m^n$. A $H$-symmetry is realised by phase space functions $s_\mu$
\begin{align}
    \{ s_\mu , s_\nu \} &= {f^\lambda}_{\mu \nu} s_\lambda , \label{eq:CartanGeometryPPGauge}\\
    \{ s_\mu , p_m \} &= {f_{\mu m }}^n p_n, \label{eq:CartanGeometryPPAction} \\
   \{ p_m , p_n \} &= 0. \label{eq:CartanGeometryPPMomenta}
\end{align}
${f^\lambda}_{\mu \nu}$ are structure constants of the Lie algebra $\mathfrak{h}$, while ${f_{\mu m}}^n$ describes the action of $H$ on the momenta. Consequentially, we impose that the momenta transform as representation of $H$. These structure constants are subjects the Jacobi identities 
\begin{equation}
    {f^\lambda}_{[\mu \nu} {f^\kappa}_{\rho] \lambda}= 0\,, 
      \qquad \text{and} \qquad
    2 {f_{[\underline{\mu} m}}^n {f_{\underline{\nu}] n}}^p = {f^\lambda}_{\mu \nu}{f_{\lambda m}}^p\,. \label{eq:JacobiIdentitiesH}
\end{equation}
The combination $J_{\mathcal{M}} = (s_\mu , p_m)$ generates the model space algebra $\mathfrak{g} = \mathfrak{h} \ltimes \mathbb{R}^{n}$ like the Euclidean or Poincar\'e algebra. This can be generalised straightforwardly to model space algebras $\mathfrak{g}$ with general structure constants ${f^{\mathcal{M}}}_{\mathcal{KL}}$, as long as there is a subalgebra $\mathfrak{h}$ and an action \eqref{eq:CartanGeometryPPAction} satisfying \eqref{eq:JacobiIdentitiesH}. Let us note that, on the phase space the Jacobi identity for $s_\mu$, $p_m$ and $x^m$, \eqref{eq:CartanGeometryPPAction} implies also an action of the gauge symmetry on the coordinates $x^m$ by 
\begin{equation}\label{eq:Salphaxl}
  \{ s_\mu , x^l\} = - {f_{\mu k}}^l x^k \,.
\end{equation}

Inspired by the discussion in the last subsection, we consider phase space functions $J_{\mathcal{A}}(x) = (j_\alpha(x) , j_a(x))$ as basis for $T_x P$. This and the model space algebra generators $J_{\mathcal{M}}$ are by the Cartan connection as $J_{\mathcal{A}} = {\theta_{\mathcal{A}}}^{\mathcal{M}} J_{\mathcal{M}}$, or equivalently
\begin{equation}
    j_\alpha = s_\alpha, \qquad j_a = {e_a}^m p_m + {\omega_a}^\mu s_\mu. \label{eq:CartanGeometryPPNewVariables}
\end{equation}
In terms of this Poisson algebra, the form the Cartan connection $\theta$ in \eqref{eq:CartanConnectionChoice} is most general that leaves \eqref{eq:CartanGeometryPPGauge} and \eqref{eq:CartanGeometryPPAction} form-invariant,
\begin{align}
    \{ j_\alpha , j_\beta \} &= {f^\gamma}_{\alpha \beta} s_\gamma \\
    \{ j_\alpha , j_b \} &= {f_{\alpha b}}^c j_c\,.
\end{align}
But in contrast to \eqref{eq:CartanGeometryPPMomenta} now the right-hand side of
\begin{equation}
    \{ j_a , j_b \} = {\Theta^{\mathcal{C}}}_{ab} J_{\mathcal{C}} = {T^c}_{ab} j_c + {R^\gamma}_{ab} j_\gamma \label{eq:CartanGeometryOrdCURRENTALGEBRA}
\end{equation}
does not vanishing any more. Instead it captures the components $\Theta_{ab}$ of the Cartan curvature \eqref{eq:CartanCurvatureNormal}. For this calculation, one needs to specify the action of the generators $s_\alpha$ on tensors -- both of $TM$ and $\mathfrak{h}$
\begin{equation}
\{ s_\alpha , T_{b ...}^{\gamma ...} \} = - \nabla_\alpha T_{b ...}^{\gamma ...} \equiv {f_{\alpha b}}^c T_{c ....}^{\gamma ...} - {f_{\alpha \beta}}^\gamma T_{b ...}^{\beta ...}\,. \label{eq:OrdHTensors} 
\end{equation}
Here, we identify it with the infinitesimal right-action of $H$, $R_h$, as required by equivariance \eqref{eq:Equivariance} above. This gives a natural notion of an infinitesimal \textit{covariant} transformation and motivates us to denote it by $\nabla_\alpha$. At the same time, we remember that in our conventions $-p_m$ acts like the partial derivative $\partial_m$ on functions that depend on the coordinates $x^n$. After contracting it with the vielbein $e_a{}^m$, the flat derivative $\partial_a = e_a{}^m \partial_m$ arises. Plugging it into the second equation of \eqref{eq:CartanGeometryPPNewVariables} and taking into account \eqref{eq:OrdHTensors}, we find the action 
\begin{equation}
  \{ j_a , T_{b ...}^{\gamma ...} \} = - \nabla_a T_{b ...}^{\gamma ...} = - (\partial_a + \omega_a^\alpha \nabla_\alpha ) T_{b ...}^{\gamma ...} \label{eq:DefNablaa}
\end{equation}
of $j_a$ on tensors. In general, transitions of indices $\mu \rightarrow \alpha$ and $m \rightarrow a$ are governed by the Kronecker symbol $\delta_\mu^\alpha$ and the vielbein ${e_a}^m$, unless stated otherwise, for example ${f_{\alpha b}}^c = \delta_\alpha^\mu {e_b}^k {e_l}^c {f_{\mu k}}^l$. The Jacobi identity requires that $\nabla_a T_{b\dots}^{\gamma\dots}$ transforms like a tensor. Hence, $\nabla_a$ is indeed a covariant derivative where $\omega$ has to transform as a connection.

\paragraph{Point Particle Dynamics.} So far, it is not obvious why the form \eqref{eq:CartanGeometryPPNewVariables} is sensible in the study of point particle dynamics. To appreciate its value, consider the standard Hamiltonian of a point particle moving in a Riemannian manifold $M$ with metric $g$:
\begin{equation}
  \mathcal{H} = \frac{1}{2} g^{mn}(x) p_m p_n.
\end{equation}
The symmetry generators $s_\mu$ are promoted to first-class constraints:
\begin{align}
    s_\mu &\approx 0,  \qquad \{ s_\alpha , s_\beta \} \approx 0 , \qquad \dot{s_\alpha} = \{ s_\alpha , \mathcal{H} \} \approx {f^{bc}}_{\alpha} j_b j_c = 0 ,
\end{align}
where the constant $\eta^{ab}$ is used to raise and lower indices $(a,b,c,...)$. Substituting to the 'Cartan basis' with \eqref{eq:CartanGeometryPPNewVariables} and assuming that the frame is chosen such that $\eta_{ab} = {e_a}^m {e_b}^n g_{mn}$, we obtain a Hamiltonian
\begin{equation}
  \mathcal{H} = \frac{1}{2} \eta^{ab} \Big(j_a - \omega_a^\alpha s_\alpha \Big) \left( j_b - \omega_b^\beta s_\beta \right) \approx \frac{1}{2} \eta^{ab} j_a j_b\,.
\end{equation}
Next, we choose the subgroup $H$ such that its action leaves $\eta_{ab}$ invariant ($\nabla_\alpha \eta_{bc} = 0$). Taking furthermore into account $\partial_m \eta_{ab} = 0$, we see that the covariant derivative introduced in \eqref{eq:DefNablaa} is metric compatible. A natural observable is the velocity of the particle
\begin{equation}
  v^a = {e_m}^a \dot{x}^m \approx \eta^{ab} j_b\,.
\end{equation}
Its equations of motion,
\begin{equation}\label{eq:GeodesicEq}
  \dot{v^a} + {\omega_{b,c}}^a v^b v^c = 0\,,
\end{equation}
produces the geodesic equation in the frame formulation with ${\omega_{b,c}}^a = {f_{\beta c}}^a \omega_b^\beta$ acting as spin connection.
For a general gauge group $H$, this will produce the auto-parallel equation
\begin{equation}
    v^a \nabla_a v^b = 0
\end{equation}
as the equation of motion with the covariant derive \eqref{eq:DefNablaa} expressed in terms of a $\mathfrak{h}$-valued 1-form connection $\omega_a^\beta$. 

\paragraph{Jacobi and Bianchi identities.} The Jacobi identities of the phase space Poisson algebra \eqref{eq:CartanGeometryOrdCURRENTALGEBRA} imply the (differential) Bianchi identities for curvature and torsion and the fact, that all components of the Cartan connection $\Theta$ transform covariantly under $H$-transformations  \eqref{eq:OrdHTensors}:
\begin{itemize}
	\item $\{ j_{[\alpha} , \{ j_\beta , j_{\gamma]} \} \} = 0, \quad \{ j_{[\alpha} , \{ j_\beta , j_{c]} \} \} = 0$:
	
	${f^\gamma}_{\alpha \beta},{f_{\alpha b}}^c$ are $H$-tensors, $\nabla_\alpha {f^\delta}_{\beta \gamma} = 0$ and $\nabla_\alpha {f_{\beta c}}^d = 0$, due to \eqref{eq:JacobiIdentitiesH}. 
	
	\item $\{ j_a , \{ j_b , s_\gamma \} \} + \text{ cyclic permutations} = 0$:
 The torsion ${T^c}_{ab}$ and curvature ${R^\gamma}_{ab}$ are $H$-tensors.
 
 \item $ \{ j_{a} , \{ j_b , j_{c} \} \} + \text{cyclic permutations} = 0$:
 
This yields the standard differential Bianchi identities for torsion ${T^c}_{ab}$ and curvature ${R^\gamma}_{ab}$:
 \begin{align}
     0 &= \left( - \nabla_{[a} {R^\delta}_{bc]} + {T^d}_{[ab} {R^\delta}_{c]d} \right) j_\delta \label{eq:OrdRicciIdentities} = \left( - \nabla_{[a} {T^d}_{bc]} + {T^e}_{[ab} {T^d}_{c]e} - {R_{[ab,c]}}^d \right) j_d 
 \end{align}
	where ${R_{ab,c}}^d = {R^\gamma}_{ab} {f_{\gamma c}}^d$.
\end{itemize}

\paragraph{Lie bracket definition of Cartan curvature.} In the Cartan geometry setting with a principal $H$-bundle $P$ and the model algebra $\mathfrak{g}$, one can define a new Lie bracket for $\mathfrak{g}$-valued fields over $P$:
\begin{equation}
    [V,W]_{L,\mathfrak{g}}^{\mathcal{M}} = V^{\mathcal{N}} D_{\mathcal{N}} W^{\mathcal{M}} - W^{\mathcal{N}} D_{\mathcal{N}} V^{\mathcal{M}} - {f^{\mathcal{M}}}_{\mathcal{KL}} V^{\mathcal{K}} W^{\mathcal{L}}, \label{eq:CartanGeometryLieBracket}
\end{equation}
with ${f^{\mathcal{M}}}_{\mathcal{KL}}$ denoting structure constants for $\mathfrak{g}$ and $D_{\mathcal{M}} = ( \nabla_\mu , \partial_m)$, where $[D_\mathcal{K},D_\mathcal{L}] = f^\mathcal{M}{}_{\mathcal{KL}} D_\mathcal{M}$. 
This is the underlying structure of the point particle phase space where such objects are realised as $V(x) = V^{\mathcal{M}}(x) J_{\mathcal{M}}$ and $W(x) = W^{\mathcal{M}}(x) J_{\mathcal{M}}$ with the Poisson bracket
\begin{equation}
  \{ V, W \} = - [ V, W ]_{L,\mathfrak{g}}^{\mathcal{M}} J_{\mathcal{M}}\,.
\end{equation}
For abelian $\mathfrak{g}/\mathfrak{h}$ with ${f^{\mathcal{M}}}_{kl}=0$, as is the case for the Poincar\'e or Euclidean algebra, this naturally defines a \textit{Lie algebroid} structure, whose anchor is the push-forward of the bundle projection $\pi$. 

In terms of this bracket, the Cartan curvature $\Theta$ \eqref{eq:CartanCurvatureNormal} arises as the bracket
\begin{equation}
    {\Theta}_{\mathcal{AB}} = - [\theta_{\mathcal{A}},\theta_{\mathcal{B}}]_{L,\mathfrak{g}} \in \mathfrak{g} \label{eq:CartanCurvatureOrdLieBracketRealisation}
\end{equation}
of the Cartan connection $\theta$ from \eqref{eq:CartanConnectionDefinition} with itself. This is not only very convenient to describe the point particle phase space and its dynamics, it also nicely generalises to generalised geometry as we discuss in section~\ref{chap:GeneralisedCartanGeometry}.

As a short summary: The coefficients in phase space Poisson algebra \eqref{eq:CartanGeometryOrdCURRENTALGEBRA} are the non-vanishing components of the (extended) Cartan curvature ${\Theta^{\mathcal{M}}}_{\mathcal{AB}}$ \eqref{eq:CartanCurvatureNormal} and have a distinct geometrical meaning: 
\begin{itemize}
  \item $\Theta^\gamma{}_{\alpha \beta  }$ $\rightarrow$ structure constants of symmetry algebra $\mathfrak{h}$\,,
  \item $\Theta^c{}_{\alpha b  }$ $\rightarrow$ action of symmetry $\mathfrak{h}$ on phase space variables\,
  \item ${\Theta^{\mathcal{M}}}_{ab}$ $\rightarrow$ curvature and torsion on $M$, as explained above.
\end{itemize}
The generalisation in section \ref{chap:GeneralisedCartanGeometry} will produce more non-vanishing components of the generalised Cartan curvature. In section \ref{chap:HigherTensors}, we will provide a complete interpretation of the additional components.

\section{Generalised Cartan geometry and the 2d $\sigma$-model current algebra} \label{chap:GeneralisedCartanGeometry}

Although not interpreted as such, what is called generalised Cartan geometry in this article has been introduced in principle in \cite{Polacek:2013nla}. There the authors used intuition from Cartan geometry in order to construct generalisations of torsion and curvature adapted to generalised geometry. This idea was explored further \cite{Butter:2022iza,Hassler:2023axp} in terms of double \cite{Siegel:1993th,Siegel:1993xq,Hull:2009mi,Hohm:2010pp} and exceptional field theory. Most of the concrete expressions in this section have been derived in these previous articles already. This section aims to show that these results fit into a natural generalisation of Cartan geometry that is presented in section \ref{chap:GeneralisedCartanGeometryDefintion}. Moreover, we discuss its realisation in terms of the string current algebra, generalising the original presentation in \cite{Polacek:2013nla} from the double Poincar\'e algebra to arbitrary model algebras.
Lastly, the passing to the extended algebra of Poláček-Siegel can be thought of strictification of the original $L_{infty}$ structure.

\subsection{Generalised Geometry} \label{chap:GeneralisedGeometryDefintion}
In order to be self-contained, let us give the most important definitions of generalised geometry and double field theory. Given a $d$-dimensional manifold $M$, generalised geometry is the geometry of a \textit{generalised tangent bundle} $(T \oplus T^\star) M$ with generalised vectors $V^M = (v^m , v_m)$. Indices $K,L,M,...$ on $(T \oplus T^\star) M$ can be raised and lowered with the \textit{invariant O$(d,d)$-metric}
\begin{equation}
	\eta_{MN} = \left( \begin{array}{cc} 0 & \delta_m^n \\ \delta_n^m & 0	\end{array} \right) \label{eq:InvariantMetric}
\end{equation}
that describes the natural pairing on $(T\oplus T^\star)M$. A generalisation of the Lie derivative on $T$ is the \textit{generalised Lie derivative} or \textit{Dorfman bracket}
\begin{align}
	\mathcal{L}_{V} W^M &= V^N\partial_N W^M - W^N \partial_N V^M + W^N \partial^M V_N \equiv [V,W]_D \label{eq:GeneralisedLieDerivative}.
\end{align}
Here, $\partial_M$ could be understood as an O$(d,d)$-covariant notation of the ordinary derivative, $\partial_M = (\partial_m , 0)$, or, in the sense of double field theory, as the derivative of an extended $2d$-dimensional geometry with extended coordinates $X^M$ being subject to the \textit{section condition}
\begin{equation}
	\eta^{MN} \partial_M \, \cdot \, \partial_N \, \cdot \, = 0, \label{eq:SectionCondition}
\end{equation}
eliminating $d$ of those coordinates. 

\subsection{Generalised Cartan geometry} \label{chap:GeneralisedCartanGeometryDefintion}
The basic definition and its underlying objects of section \ref{chap:CartanGeometryOrdinary} generalise naturally to generalised geometry. The key to this generalisation is that a generalised Cartan connection links the \textit{generalised tangent bundle} $(T\oplus T^\star)P$ with $2d'$-dimensional $\eta$-compatible Lie \textit{algebra} (instead of a Lie algebra). The definition of the generalised Cartan curvature is possible due to a Dorfman bracket on $(T \oplus T^\star)P$, which has properties similar to the Lie bracket \eqref{eq:CartanGeometryLieBracket} in section~\ref{chap:PointParticle}. 

In the following, we will go through the fundamental objects and definitions from the last section and show how they are modified and extended in the framework of generalised geometry. We begin with
\begin{enumerate}
    \item The underlying groups:
    \begin{itemize}
        \item A Lie group $H$, called \textit{gauge group}, with Lie algebra $\mathfrak{h}$ and its associated principal bundle $P$, $\pi: \ P \rightarrow M$.
        
        In contrast to ordinary Cartan geometry, here the prototypical example will be the double Lorentz group\footnote{For brevity of notation, we will write the double Lorentz group typically as O$(d)\times$O$(d)$, but any other signature works as well.} O$(d) \times$O$(d)$. This is the a generalisation of the Lorentz group that has a natural action on the generalised tangent bundle $(T\oplus T^\star)M$. It is the reduced structure group after fixing a generalised metric which is stabilised by the action of this group.
        
      \item The \textit{model algebra} is a $2d'$-dimensional quasi-Lie bialgebra $\mathfrak{d}$, where $d' = d + \text{dim}(\mathfrak{h})$ with a isotropic subalgebra $\mathfrak{h}$. A quasi-Lie bialgebra is a Lie algebra with a \textit{non-generate, ad-invariant symmetric pairing} $\eta$, such that as vector space it decomposes into a $d'$-dimensional vector space and its dual. $\eta$ captures the natural pairing between these two dual vectors spaces and therefore corresponds to the invariant O$(d',d')$-metric we already encounter in \eqref{eq:InvariantMetric}. 
        
        Moreover, for the generalised Cartan geometry setting, $\mathfrak{d}$ has to contain $\mathfrak{h}$ as a (non-maximal) \textit{isotropic subalgebra} with 
        \begin{equation}
          [ \mathfrak{h} , \mathfrak{h} ] \subset \mathfrak{h}, \qquad \text{and} \qquad \eta(\mathfrak{h},\mathfrak{h}) = 0.
        \end{equation}
As a consequence, one can decompose $\mathfrak{d}$ as a direct sum of vector spaces, but \textit{not} of Lie algebras, as
\begin{equation}
	\mathfrak{d} = \mathfrak{h} \oplus \tilde{\mathfrak{d}} \oplus \mathfrak{h}^\star
\end{equation}
for some $2d$-dimensional vector space $\tilde{\mathfrak{d}}$. In indices, this corresponds to 
        \begin{equation}
	\eta_{\mathcal{M}\mathcal{N}} = \left( \begin{array}{ccc} 0 & 0 & \delta_\mu^\nu \\ 0 & \eta_{MN} & 0 \\ \delta_\nu^\mu & 0 & 0	\end{array} \right) \label{eq:GeneralisedMetricExtended}
\end{equation}
where $\mathcal{K},\mathcal{L},\mathcal{M},...$ are indices on $\mathfrak{d}$, that decompose as $T_\mathcal{K} = (t_\kappa , t_K , t^\kappa)$, and $\kappa,\lambda,\mu,...$ are indices on $\mathfrak{h}$. We write the generators and structure constants of $\mathfrak{d}$
\begin{equation}
	[ T_{\mathcal{K}} , T_{\mathcal{L}} ] = {f^{\mathcal{M}}}_{\mathcal{K} \mathcal{L}} T_{\mathcal{M}}.
\end{equation}

        In this article, we consider two cases for model algebras $\mathfrak{d}$:
\begin{itemize}
	\item Let $H$ be a Lie group with an action on $(T\oplus T^\star)M$, the prototypical example being the double Lorentz group O$(d)\times$O$(d)$. This corresponds to a Lie algebra $\mathfrak{h} \ltimes \mathbb{R}^{d,d}$. This is, by construction, \textit{not} a quasi-Lie bialgebra by itself -- the bilinear form is degenerate. But it has a canonical extension to a quasi-Lie bialgebra $\mathfrak{d}$. This construction is presented in section \ref{chap:GeneralisedCartanGeometryModelBialgebra}, generalising \cite{Polacek:2013nla}. For $\mathfrak{h}$ being the double Lorentz algebra, we call this the \textit{double Poincar\'e} algebra.

	\item Background geometries that are interesting for generalised dualities are the motivation behind \cite{Butter:2022iza} and correspond to dressing cosets $H/D\backslash F$ with some underlying double Lie group $D$. Then the underlying model algebra $\mathfrak{d}$ is given as the Lie algebra to $D$. Its necessary assumptions here are, in general, smaller 
\end{itemize}
    \end{itemize}
    
    \item A \textit{generalised Cartan connection} $\theta$ is an isomorphism
    \begin{equation}
        \theta(p): \ (T \oplus T^\star)_p P \rightarrow \mathfrak{d} \label{eq:GeneralisedCartanConnectionDefinition}
    \end{equation}
    that identifies the canonical pairing on $(T\oplus T^\star)P$ with the one of $\mathfrak{d}$, implying $\theta \in$O$(d',d')$. This definition is the central difference to ordinary Cartan geometry. Compared to standard Cartan geometry case, it only defines
    \begin{equation}
      (T \oplus T^\star)_{\pi(p)} \simeq \mathfrak{\tilde{d}} = \mathfrak{d}/(\mathfrak{h} + \mathfrak{h}^\star)\,,
    \end{equation}
    necessitating a choice of dual algebra $\mathfrak{h}^\star$ that has no obvious geometric meaning but rather is a redundancy of our description. Hence, in generalised Cartan geometry, the generalised Cartan connection contains an ambiguity depending on the choice of $\mathfrak{h}^\star \subset \mathfrak{d}$, that will correspond to a new gauge field as component of $\theta$. Note that $\mathfrak{h}^\star$ is a choice of dual vector space to $\mathfrak{h}$ but there are no restrictions on its algebraic structure.
    
    The further standard assumptions on $\theta$ on Cartan geometry are the same as in section \ref{chap:CartanGeometryOrdinary}, mimicking \eqref{eq:NormalCartanIdentity} and \eqref{eq:Equivariance}:
 \begin{itemize}
        \item Left-invariant vector fields of $TH$ are identified with the subalgebra $\mathfrak{h} \subset \mathfrak{d}$, via the absolute parallelism. We denote them by $X_\xi$, with $\xi\in\mathfrak{h}$, and the Cartan connection has to satisfy
          \begin{equation}\label{eq:CartanIdentity}
            \theta(X_\xi) = \xi\,.
          \end{equation}
        \item \textit{Equivariance}. The right-action $R_h$ of $h \in H$ has to match the adjoint action of $h^{-1}$, through
        \begin{equation}
            R^\star_h \theta = \mathrm{Ad}_{h^{-1}} \theta\,. \label{eq:EquivarianceGeneralised}
        \end{equation}
        Again, we will interpret this as gauge transformations for the components of $\theta$ \eqref{eq:ConnectionGaugeRegular}.
    \end{itemize}
    When we take into account that $\theta$ has to be an O$(d',d')$-transformation and is further constraint by \eqref{eq:CartanIdentity}, we are left with the most general expression
    \begin{equation} 
	{\theta_\mathcal{A}}^{\mathcal{M}}(X) = \left( \begin{array}{ccc} \delta_\alpha^\mu & 0 & 0 \\ \Omega_A^\mu(X) & {E_A}^M(X) & 0 \\ \rho^{\alpha \mu}(X) - \frac{1}{2} \Omega^{B\alpha}(X) \Omega_B^\mu(X) & - \Omega^{\alpha}_A(X) {E^A}_M(X) & \delta^\alpha_\mu
	\end{array} \right) . \label{eq:GeneralisedCartanConnectionChoice}
\end{equation}
    for generalised Cartan connection where $\mathcal{A}=({}_\alpha,A,{}^\alpha),\mathcal{B},...$ denote indices on 
    \begin{equation}
    (T \oplus T^\star)_p P \simeq (\mathfrak{h} \oplus \mathfrak{h}^\star) \oplus (T \oplus T^\star)_{\pi(p)} M .
    \end{equation}
    The non-trivial part of the Cartan connection decomposes into a connection $\Omega_M \in \mathfrak{h}$ on $M$, the generalised frame field ${E_A}^M$ and the field $\rho \in \mathfrak{h} \wedge \mathfrak{h}$. Latter is novel compared to standard Cartan geometry \eqref{eq:CartanConnectionChoice} and the common analysis in generalised geometry or double field theory. It is interpreted as the gauge field that is associated to the freedom of choosing a dual vector space $\mathfrak{h}^\star \subset \mathfrak{d}$ to $\mathfrak{h}$. Due to its first appearance in \cite{Polacek:2013nla}, we call it \textit{Poláček-Siegel field} in the following. In case that $H$ is the double Lorentz group, $\Omega$ can be interpreted as generalised affine connection, as will be demonstrated in section \ref{chap:GeneralisedMetric}.

	\item For $\mathfrak{d}$-valued fields $V,W$ over $M$ a \textit{$\mathfrak{d}$-twisted Dorfman bracket} is defined in analogy to \eqref{eq:CartanGeometryLieBracket} as
	\begin{align}
		 [V,W]_{D,\mathfrak{d}}^{\mathcal{M}} \label{eq:GeneralisedCartanGeometryDorfmannBracket} 
		&= V^{\mathcal{N}} \mathcal{D}_{\mathcal{N}} W^{\mathcal{M}} - W^{\mathcal{N}} \mathcal{D}_{\mathcal{N}} V^{\mathcal{M}} + W^{\mathcal{N}} \mathcal{D}^{\mathcal{M}} V_{\mathcal{N}} - {f^{\mathcal{M}}}_{\mathcal{K}\mathcal{L}} V^{\mathcal{K}} W^{\mathcal{L}}
	\end{align}	 
  where $\mathcal{D}_{\mathcal{M}} = (\nabla_\mu , D_M , \tilde{\partial}^\mu) = (\nabla_\mu , D_M , 0)$\footnote{In the extended space \cite{Butter:2022iza}, the section condition on the coordinates $y$ of $H$ is solved by eliminating any dependence on the dual coordinates $\tilde{y}$. Therefore, $\tilde{\partial}^\mu \, \cdot \, = 0$ holds.} and $\partial_M$ is subject to the section condition \eqref{eq:SectionCondition}. Again, $\mathcal{D}_\mathcal{K}$ are not flat derivatives but correspond to a non-holonomic basis on $(T\oplus T^\star) P$. In particular,
    \begin{equation}
    		[\mathcal{D}_\mathcal{K} , \mathcal{D}_\mathcal{L}] = {f^\mathcal{M}}_{KL} \mathcal{D}_\mathcal{M}\,.
    \end{equation}
    Similar to the ordinary Cartan geometry, for the case where $\mathfrak{d}$ is a generalisation of the Poincar\'e algebra, one can assume that the corresponding derivatives commute 
\begin{equation}
  [D_M , D_N] = f_{\alpha MN} \tilde{\partial}^\alpha = 0, \label{eq:GeneralisedPoincareType}
\end{equation}   
such that they can be identify with the flat derivatives $\partial_M$ related to the extended coordinates $X^M$ on $M$. This also means that, for the model algebras of what we call being 'generalised Poincar\'e type', $\mathfrak{d}$-valued fields over $M$ together with $[ \ , \ ]_{D,\mathfrak{d}}$ define a (non-exact) \textit{Courant algebroid} over $M$.

    \item The \textit{generalised Cartan curvature} $\Theta$ for the generalised Cartan connection $\theta$ is
        \begin{equation}
            \Theta_{\mathcal{AB}} = - [\theta_{\mathcal{A}} , \theta_{\mathcal{B}} ]_{D,\mathfrak{d}} \in \mathfrak{d}. \label{eq:GeneralisedCartanCurvatureNormal}
        \end{equation}
        The interpretation of components of $\Theta$ as torsions -- components whose fixing leads to algebraic constraints to components of the connection $\theta$ -- or as curvatures -- containing genuine geometric information will be discussed in section \ref{chap:HigherTensors}.
\end{enumerate}

\paragraph{Generalised Cartan Geometry from Extended Double Field Theory.} In the spirit of double field theory one can consider an extended space $\mathcal{M}$ associated to $(T \oplus T^\star)P$. This space is $2d'$-dimensional with coordinates $Y^{\mathcal{M}} = (y^\alpha , X^M , y_\alpha)$ and is called \textit{megaspace} in \cite{Butter:2022iza}. $\mathcal{M}$ has a canonical O$(d',d')$-metric of the form \eqref{eq:GeneralisedMetricExtended}, an associated generalised Lie derivative
\begin{equation}
	[V,W]_D^{(ext)} = V^{\mathcal{N}} \partial_{\mathcal{N}} W^{\mathcal{M}} - W^{\mathcal{N}} \partial_{\mathcal{N}} V^{\mathcal{M}} + W^{\mathcal{N}} \partial^{\mathcal{M}} V_{\mathcal{N}}\,,  \label{eq:MegaExtendedGeneralisedLieDerivative}
\end{equation}

The drinfeld bracket on $(T+T^*)M$ can be obtained by considering an uplift of the following natural projections of the generalized bundles 
\begin{equation} \label{proj}
     (T+T^*) P \rightarrow (T+T^*)M
\end{equation}
the uplift $A$ takes a section of $(T+T^*)M$ and sends the $\partial^{\mathcal{M}} V_{\mathcal{N}}$ to a generator of $\mathcal{h}$ alongside with the identity on $(T+T^*)M$ part of $(T+T^*)P$. Then one can write 
\begin{equation}
    [V,W]_D^M = \pi [A(V),A(W)]^P
\end{equation}
where the $\pi$ is a bundle projection \eqref{proj} and $[\, ,\,]^P$ is a regular Lie bracket on $P$.

and the section condition \eqref{eq:SectionCondition}. This ensures that physical quantities only depend on $y^\alpha$ (fibre coordinates on $P$) and a section of $X^M$ as in ordinary double field theory. In generalised Cartan geometry the possible dependence of physical quantities on the coordinates $y$ is dropped. 

But the two approaches do not contradict each other. They are equivalent. As we discussed in section~\ref{chap:CartanGeometryOrdinary}, an appropriate twist allows to derive the fundamental relations \eqref{eq:NormalCartanIdentity} and \eqref{eq:Equivariance} from an extended space. Here, we follow the same route to show that generalised Cartan geometry can be reproduced from an \textit{extended generalised vielbein} in extended double field theory. Following our insights from \eqref{eq:twistForCartan}, we again parameterise it as
\begin{equation}
  \widehat{\theta} = \widetilde{V}_{\mathfrak{d}} \overline{\theta} \widetilde{M}_{\mathfrak{d}}\,, \label{eq:MegaExtendedVielbein}
\end{equation}
where the definition of $\widetilde{M}_{\mathfrak{d}}$ is the same as in \eqref{eq:DefTildeMg} just for generators in the extended Lie algebra $\mathfrak{d}$ instead of $\mathfrak{g}$. As expected, $\overline{\theta}$ is also restricted to the form given in \eqref{eq:GeneralisedCartanConnectionChoice}. The only deviation from the previous discussion is $\widetilde{V}_{\mathfrak{d}}$. It has to be an element of O$(d',d')$, which is achieved by modifying \eqref{eq:DefTildeVg} according to
\begin{equation}
  \widetilde{V}_{\mathfrak{d}} = \widetilde{v} \oplus \mathbf{1}_{\mathfrak{d}/\tilde{\mathfrak{d}}} \oplus \widetilde{v}^{-T}
\end{equation}
while preserving the crucial property
\begin{equation}
  R_h^* \widetilde{V}_{\mathfrak{d}} = \widetilde{V}_{\mathfrak{d}}\,.
\end{equation}
With these adapted definitions, one recovers the analog versions
\begin{equation}
  R^*_h \widehat{\theta} = R_h \widetilde{M}_{\mathfrak{d}} R_h \overline{\theta} R_h \widetilde{V}_{\mathfrak{d}} = \mathrm{Ad}_{h^{-1}} \widehat{\theta}
\end{equation}
of \eqref{eq:EquivarianceExplicit} and \eqref{eq:CartanIdentityExplicit}.  From the perspective of Cartan geometry, $\widetilde{M}_{\mathfrak{d}}(y)$ defines the model algebra $\mathfrak{d}$ and $\overline{\theta}(X)$ plays the role of the generalised Cartan connection. Furthermore, the twist $M_{\mathfrak{d}}(y)$ mediates between $[V,W]^{(ext)}_D$ in \eqref{eq:MegaExtendedGeneralisedLieDerivative} and $[V,W]_{D,\mathfrak{d}}$ in \eqref{eq:GeneralisedCartanGeometryDorfmannBracket} through
\begin{equation}
  \Theta_{\mathcal{A} \mathcal{B}} = - [\theta_{\mathcal{A}}, \theta_{\mathcal{B}} ]_{D,\mathfrak{d}} = - \widetilde{M}^{-1} (\widetilde{M}^{-1})_{\mathcal{A}}{}^{\mathcal{C}} (\widetilde{M}^{-1})_{\mathcal{B}}{}^{\mathcal{D}} [\widehat{\theta}_{\mathcal{C}} , \widehat{\theta}_{\mathcal{D}} ]_D^{(ext)}
\end{equation}
which provides an alternative way to compute the generalised Cartan curvature \eqref{eq:GeneralisedCartanCurvatureNormal}. Hence, in the language of generalised geometry the generalised Cartan curvature is nothing else than the \textit{generalised flux} or \textit{generalised torsion} on the megaspace for the extended generalised vielbein \eqref{eq:MegaExtendedVielbein}. Note that the ansatz for the latter breaks the O$(d',d')$-covariance to O$(d,d) \times$O$(n,n)$ due to the separation of coordinates $Y = (y,X)$. For objects (other than this frame \eqref{eq:MegaExtendedVielbein}) we assume a section: $\mathcal{D}_\mathcal{M} = (\nabla_\mu , \partial_M , 0)$, where $\partial_M$ is subject to the ordinary section condition \eqref{eq:SectionCondition}.

\paragraph{Construction from the current algebra.} Originally, Poláček and Siegel \cite{Polacek:2013nla} derived a realisation of the above setting on the phase space Poisson algebra of the bosonic string\footnote{or any bosonic 2d non-linear $\sigma$-model for that matter} -- the classical \textit{string current algebra}. The phase space variables of a string moving in a $d$-dimensional manifold $M$ can be arranged in terms of an O$(d,d)$-vector $P_M(\sigma) = (p_m(\sigma) , \partial x^m(\sigma) )$, where $\sigma$ is the spatial world-sheet coordinate and $\partial \equiv \partial_\sigma$. As discussed from many angles in the literature \cite{Duff:1989tf,Tseytlin:1990nb,Tseytlin:1990va,Siegel:1993bj,Siegel:1993th,Klimcik:1995ux,Sfetsos:1999cc,Blair:2014kla,Demulder:2018lmj,Osten:2019ayq,Hassler:2020xyj,Borsato:2021gma,Borsato:2021vfy}, the \textit{current algebra}
\begin{equation}
    \{ P_M (\sigma_1) , P_N (\sigma_2) \} = \eta_{MN} \delta^\prime (\sigma_1-\sigma_2) \label{eq:CurrentAlgebraFlatNotExtended}
\end{equation}
realises the generalised Lie derivative as $\{ W , V \} = [V,W]_D$ up to boundary terms \cite{Osten:2019ayq}. 
This framework can be extended to the heterotic string \cite{Hatsuda:2022zpi,Osten:2023cza}, and brane $\sigma$-models in supergravity \cite{Osten:2021fil,Osten:2024mjt}.

As in the point particle case, a (gauge) symmetry $\mathfrak{h}$ can be introduced by phase space functions under the constraint $s_\mu \approx 0$. It is this realisation, inspired by \cite{Polacek:2013nla}, that we will present in the next subsection. 

\subsection{Current algebra realisation of the model algebra \texorpdfstring{$\mathfrak{d}$}{d}} \label{chap:GeneralisedCartanGeometryModelBialgebra}
We will differentiate between two cases for the model algebra $\mathfrak{d}$. This is necessary because given a Lie algebra $\mathfrak{h}$ and its action on the generalised tangent bundle does \textit{not} yet define a quasi-Lie bialgebra $\mathfrak{d}$ that could serve as a model algebra.
\begin{itemize}
	\item In case of $\mathfrak{h}$ being the double Lorentz algebra, a construction to obtain a quasi-Lie bialgebra -- in particular with a non-generate bilinearform -- was given in \cite{Polacek:2013nla}. We discuss generalisations of this approach, which we refer to as \textit{'generalised Poincar\'e'-type}, below. 
  \item Alternatively, a generic quasi-Lie bialgebra $\mathfrak{d}$ with an isotropic subalgebra $\mathfrak{h}$, could be given as input. This for example happens for target spaces of the form $H\backslash D /G$. They are called generalised cosets \cite{Demulder:2019vvh} and have proven central for generalised $T$-dualities and the related consistent truncations \cite{Butter:2022iza}. Here $\mathfrak{d}$ is known from the beginning and does not need to be constructed. Model algebras of the generalised Poincar\'e-type are a special case.
\end{itemize}
We will present a realisation as current algebra for the second, general case after reviewing the structure of generalised Poincar\'e-type model algebra. 

\paragraph{Model algebra of the 'generalised Poincar\'e'-type.}
On the string phase space, the gauge symmetry $\mathfrak{h}$ is introduced as (local) action of generating currents $s_\mu (\sigma)$ on the generalised tangent bundle together with \eqref{eq:CurrentAlgebraFlatNotExtended}.
\begin{align}
    \{s_\mu(\sigma_1) , s_\nu(\sigma_2)\} &= {f_{\mu \nu}}^\lambda s_\lambda(\sigma_1) \delta(\sigma_1 - \sigma_2), \label{eq:GeneralisedCartanGeometryGauge} \\
    \{ s_\lambda(\sigma_1) , P_M(\sigma_2) \} &= {f_{\lambda M}}^N P_N(\sigma_1) \delta(\sigma_1 - \sigma_2), \label{eq:GeneralisedCartanGeometryAction}
\end{align}
where ${f_{\mu \mu}}^\lambda$ are structure constants of $\mathfrak{h}$ and ${f_{\lambda M}}^N$ describes the action of $\mathfrak{h}$ on the generalised tangent bundle. They satisfy the identities \begin{equation}
    {f^\lambda}_{[\mu \nu} {f^\kappa}_{\rho] \lambda}= 0, \qquad 2 {f_{[\underline{\mu} M}}^N {f_{\underline{\nu}] N}}^P = {f^\lambda}_{\mu \nu}{f_{\lambda M}}^P . \label{eq:JacobiIdentitiesH2}
\end{equation}
Such an algebra was introduced in \cite{Polacek:2013nla} for $\mathfrak{h}$ being the double Lorentz algebra. The construction is totally general as long as the form of the action on the generalised tangent bundle is of the form \eqref{eq:GeneralisedCartanGeometryAction}. This current algebra has two problems:
\begin{itemize}
	\item Its zero mode subalgebra, which will correspond to the model algebra $\mathfrak{d}$, is not a quasi-Lie bialgebra. As zero modes we understand: $s_\mu = \int \mathrm{d} \sigma \ s_\mu(\sigma)$ and correspondingly for $P_M(\sigma)$. 
	\item Moreover, the Jacobi identity of the Poisson brackets is violated as one can see for example from
\begin{align*}
	&{} \quad \{ s_\mu (\sigma_1) , \{ P_K (\sigma_2) , P_L (\sigma_3) \} \} -
	 \{ \{ s_\mu (\sigma_1) , P_K (\sigma_2) \} , P_L (\sigma_3) \} - \{ P_K (\sigma_2) , \{ s_\mu (\sigma_1) , P_L (\sigma_3) \} \} \\
	 &= f_{\mu MN} \left( \delta(\sigma_1 - \sigma_3) \delta^\prime (\sigma_2 - \sigma_3) - \delta(\sigma_1 - \sigma_2) \delta^\prime(\sigma_1 - \sigma_3) \right) \neq 0.
\end{align*}
	\end{itemize}
A simultaneous solution for both problems \cite{Polacek:2013nla} is to introduce dual gauge symmetry generators $\Sigma^\mu$ to complement the generators $s_\mu$. They come with the Poisson brackets
\begin{align}
    \{ P_M (\sigma_1) , P_N (\sigma_2) \} &= \eta_{MN} \delta^\prime (\sigma_1-\sigma_2)  + f_{\mu MN} \Sigma^\mu (\sigma_1) \delta(\sigma_1 - \sigma_2) \nonumber \\
    \{ s_\mu(\sigma_1), \Sigma^\nu(\sigma_2) \} &= \delta_\mu^\nu \delta^\prime(\sigma_1 - \sigma_2)  + {f_{\kappa \mu}}^\nu \Sigma^\kappa(\sigma_1)  \delta(\sigma_1- \sigma_2) \nonumber \\
    \{ \Sigma^\mu(\sigma_1) , \Sigma^\nu(\sigma_2) \} &= 0 , \nonumber \\
    \{ P_M(\sigma_1) , \Sigma^\alpha(\sigma_2) \} &= 0 \label{eq:CurrentAlgebraFlat}
\end{align}
in addition to \eqref{eq:GeneralisedCartanGeometryGauge} and \eqref{eq:GeneralisedCartanGeometryAction}. Starting from them, one can verify that the zero mode algebra generated by $s_\mu,P_M,\Sigma^\mu$ 
\begin{align} \label{zeromode}
  \{s_\mu, s_\nu\} &= {f_{\mu \nu}}^\kappa s_\kappa, & \{ s_\mu , P_M \} &= {f_{\mu M}}^N P_N\,, \nonumber \\
  \{ P_M, P_N \} &=  f_{\mu MN} \Sigma^\mu, & \{ s_\mu, \Sigma^\nu \} &= {f_{\kappa \mu}}^\nu \Sigma^\kappa\,, \nonumber \\
  \{ \Sigma^\mu , \Sigma^\nu \} &= 0 , & \{ P_M , \Sigma^\alpha \} &= 0
\end{align}
together with pairing \eqref{eq:GeneralisedMetricExtended} is indeed now a quasi-Lie bialgebra. Also the Jacobi identity of the current algebra \eqref{eq:CurrentAlgebraFlat} is now satisfied.

We name model algebras of this type as being of \textit{generalised Poincar\'e type} -- as explained above, as an algebra describing a (quasi-) flat model space (through \eqref{eq:GeneralisedPoincareType}), with an action of the gauge group $H$ on this space.

\paragraph{Extension of generalised Poincar\'e by ${f_\lambda}^{\mu \nu}$ or ${f}^{\lambda \mu \nu}$.}
It might seem that one could easily generalise the algebra $\mathfrak{h}^\star$ generated by the $\Sigma^\alpha$ to be non-abelian or closing into $\mathfrak{h}$. But it turns out that the form \eqref{eq:CurrentAlgebraFlat} is quite unique when the action of $\mathfrak{h}$ is non-trivial. There can be none of the underlined terms in the current algebra
\begin{align*}
\{ s_\mu(\sigma_1), \Sigma^\kappa(\sigma_2) \}
& = \delta_\mu^\nu \delta^\prime(\sigma_1 - \sigma_2) + \left( {f_{\lambda \mu}}^\kappa \Sigma^\lambda(\sigma_1) + \underline{ {f_\mu}^{\kappa \lambda} }s_\lambda (\sigma_1) \right) \delta(\sigma_1- \sigma_2) \\
    \{ \Sigma^\mu(\sigma_1) , \Sigma^\nu(\sigma_2) \} &=  \underline{{f_\lambda}^{\mu \nu}} \Sigma^\lambda (\sigma_1)  \delta(\sigma_1- \sigma_2) +  \underline{ {f}^{\lambda \mu \nu}} s_\lambda (\sigma_1)  \delta(\sigma_1- \sigma_2)  \\
\end{align*}
due ${f^\mu}_{MN} = 0$ which guarantees a flat model space. In principle, both might be allowed in the cases where
\begin{equation}
	{f_\lambda}^{\mu \nu} {f_{\nu M}}^N = 0, \qquad f^{\lambda \mu \nu} {f_{\nu M}}^N = 0.
\end{equation}
This would mean on the fixed points of the action of $\mathfrak{h}$ they could be chosen to be non-vanishing. Such constants will be set to be vanishing in the following. Conceptually, this is might also be preferable, as then the whole current algebra of the $(S_\mu , P_M , \Sigma^\mu)$ is defined only be the action of $F$ (resp. the generators $S_\mu$) and no additional input is given.

\paragraph{General model algebra.} Given a $2d'$-dimensional quasi-Lie bialgebra $\mathfrak{d}$ with $n$-dimensional isotropic subalgebra $\mathfrak{h}$ and structure constants ${f^{\mathcal{K}}}_{\mathcal{M}\mathcal{N}}$, there is a natural current algebra realisation generalising the previous 'generalised Poincar\'e'-type model algebras. Let $J_{\mathcal{M}} = (s_\mu , P_M , \Sigma^\mu)$, then 
\begin{equation}
	\{ J_{\mathcal{M}} (\sigma_1) , J_{\mathcal{N}} (\sigma_2) \}= \eta_{\mathcal{MN}} \delta^\prime(\sigma_1 - \sigma_2) + {f^{\mathcal{K}}}_{\mathcal{M}\mathcal{N}} J_{\mathcal{K}}. \label{eq:GeneralisedCartanCurrentAlgebraFlatGeneral}
\end{equation}
This gives the current algebra equivalent of the the $\mathfrak{d}$-twisted Dorfman bracket \eqref{eq:GeneralisedCartanGeometryDorfmannBracket}
\begin{equation}
	\{V , W\} = - [V,W]_{D,\mathfrak{d}}
\end{equation}
with $V = \int \mathrm{d} \sigma \ V^{\mathcal{M}} J_{\mathcal{M}}$.

Let us give one prototypical example besides the 'generalised Poincar\'e'-type which is characterised by the only non-vanishing components of ${f^{\mathcal{K}}}_{\mathcal{M}\mathcal{N}}$ being ${f^\lambda}_{\mu\nu}$ and ${f_{\lambda M}}^N$. Consider a reductive coset space $G/H$ for a $D$-dimensional Lie group $G$ with Lie algebra $\mathfrak{g}$, then there is always the so-called \textit{semi-abelian Lie bialgebra}:
\begin{equation}
\mathfrak{d} = \mathfrak{g} \ltimes (\mathfrak{u}(1))^D \nonumber ,
\end{equation}
which is characterised by the non-vanishing structure constants ${f^\lambda}_{\mu\nu}, {f^m}_{kl}, {f^\mu}_{kl}, {f_{\lambda m}}^n$. This structure appears in the study of non-abelian T-duality \cite{Klimcik:1995ux}. It is also known that this is generally not the only Lie bialgebra, or weaker: Manin pair, one can build from $\mathfrak{g}$ \cite{Vicedo:2015pna}.

\paragraph{Action on functions and tensors.} In order to proceed, the action of $(s_\mu , P_M , \Sigma^\mu)$ on functions of the coordinates should be specified. By construction
\begin{equation}
	\{P_M(\sigma_1) , f(X(\sigma_2)) \} = -\partial_M f(X(\sigma_1)) \delta(\sigma_1 - \sigma_2)
\end{equation}
holds. Imposing the Jacobi identity between (generalised) coordinates $X^M$ and the other current algebra generators we find
\begin{equation}\label{eq:actionOnCoordinates}
  \{ s_\kappa (\sigma_1) , X^M(\sigma_2) \} = - {f_{\kappa L}}^M X^L 
\end{equation}
from Jac$(s,P,X)$ and
\begin{equation}
  \{ \Sigma^\kappa (\sigma_1) , X^M(\sigma_2) \} = 0
\end{equation}
from Jac$(s,\Sigma,X)$. We can now follow the same steps as for the point particle in section~\ref{chap:PointParticle}. We introduce
\begin{equation}
  \{ s_\kappa , \, \cdot \, \} = - \nabla_\kappa \, \cdot \,
\end{equation}
under with tensors transform as in \eqref{eq:OrdHTensors} but with adapted indices, giving rise to
\begin{equation}
  \{ s_\alpha, T^{\gamma\dots}_{\beta\dots\,D\dots} \} = - \nabla_\alpha T^{\gamma\dots}_{\beta\dots\, D\dots} \equiv -f_{\alpha\beta}{}^\delta T^{\gamma\dots}_{\delta\dots\,D\dots} + f_{\alpha\delta}{}^\gamma T^{\delta\dots}_{\beta\dots\,D\dots} + f_{\alpha D}{}^E T^{\gamma\dots}_{\beta\dots\,E\dots}\,.
\end{equation}
Remarkably, lifting the restriction on the model algebra $\mathfrak{d}$ being of generalised Poincar\'e-type permits a richer structure of covariant transformations. When requiring equivariance \eqref{eq:EquivarianceGeneralised}, it turns out that the covariant derivatives of objects in the generalised Cartan connection have to be modified by translation terms -- schematically: $\nabla T = f T + \text{translation terms}$ -- in the case a generic model algebra $\mathfrak{d}$, i.e. not of generalised Poincar\'e type. will be related to components of the structure constants. We will discuss the explicit form of the covariant transformations in the next subsection. 


\subsection{The generalised Cartan curvature from the current algebra} \label{chap:GeneralisedCartanGeometryCurvature}
In analogy with the point particle, the generalised Cartan connection defines the most general change of variables on the string phase space, $
J_\mathcal{A}= (J_\alpha , J_A , J^\alpha) = {\theta_\mathcal{A}}^{\mathcal{M}} J_\mathcal{M}$, that keeps the form of \eqref{eq:GeneralisedCartanGeometryGauge} and \eqref{eq:GeneralisedCartanGeometryAction} form-invariant. 
\begin{equation}
    \{ J_\mathcal{A}(\sigma_1) , J_\mathcal{B}(\sigma_2) \} = \eta_{\mathcal{AB}} \delta^\prime(\sigma_1 - \sigma_2) + {\Theta^\mathcal{C}}_{\mathcal{AB}}(\sigma_1) J_\mathcal{C}\delta(\sigma_1 - \sigma_2) \label{eq:GeneralisedCartanCurvatureCurrent}
\end{equation}
The only difference to the calculation via the definition \eqref{eq:GeneralisedCartanCurvatureNormal} is that it is natural to obtain the components of $\Theta$ as ${\Theta^\mathcal{C}}_{\mathcal{AB}} = {\theta_\mathcal{M}}^\mathcal{C}(X) {\Theta^\mathcal{M}}_{\mathcal{AB}}$ in the current algebra. As the generalised Cartan connection defines an isomorphism, both results are equivalent.

Again we differentiate between $\mathfrak{d}$ being of 'generalised Poincar\'e'-type, or a generic quasi-Lie bialgebra with isotropic subalgebra $\mathfrak{h}$. Former reproduces the standard generalised curvature and torsion of generalised geometry and double field theory, as will be shown in section \ref{chap:GeneralisedMetric}.

\paragraph{'Generalised Poincar\'e'-type.} All components of the generalised Cartan curvature can be derived from the current algebra. They take the following explicit form:
\begin{align} \label{curvature}
	{\Theta_{\alpha \beta}}^\gamma &= {f_{\alpha \beta}}^\gamma \nonumber \\
	{\Theta_{\alpha B}}^C &= {f_{\alpha B}}^C - \nabla_\alpha {E_B}^M {E_M}^C \nonumber \\
	{\Theta_{\gamma}}^{\alpha \beta} &= 0 = {\Theta_{\alpha B}}^\gamma
\end{align}
with ${f_{\alpha B}}^C = {f_{\alpha M}}^N {E^M}_B {E_N}^C$ and similarly for other objects. Moreover, we defined $F_{ABC} = 3 {E_{[A}}^M \partial_B E_{C]M}$, ${\Omega_{A,B}}^C = {\Omega_A^\alpha} {\Theta_{\alpha B}}^C$ and $\tilde{\rho}^{\alpha \beta} = \rho^{\alpha \beta} - \frac{1}{2} \Omega^{\alpha}_A \Omega^{\beta A}$. In addition, the coset constraints
\begin{align}\label{eq:nablaalphas}
	\nabla_\alpha {E_B}^M &\equiv - {\Theta_{\alpha B}}^C {E_C}^M + {f_{\alpha L}}^M {E_B}^L, \\
	\nabla_\alpha \Omega_B^\gamma &\equiv - {\Theta_{\alpha B}}^C \Omega_C^\gamma + {f_{\alpha \beta }}^\gamma \Omega_B^\beta, \label{eq:ConnectionGaugeRegular}\\
	\nabla_\alpha \rho^{\beta\gamma} &\equiv - 2 {f^{[\beta}}_{\alpha \delta} \rho^{\gamma] \delta }, \nonumber
\end{align}
are imposed, such that \ $\rho$, $\Omega$ and $E$ transform under $H$-gauge transformations according to equivariance \eqref{eq:EquivarianceGeneralised} of the generalised Cartan connection. This fixes $f_{\alpha B}{}^\gamma = 0$, $f_{\alpha}{}^{\beta\gamma}=0$ and completes all data necessary the model algebra $\mathfrak{d}$.

The remaining components are denoted by the bold quantities 
\begin{align} 
	{\mathbf{\Theta}}_{ABC} &= - 3 {E_{[C}}^M \partial_A E_{B]M} + 3 \Theta_{\gamma [AB} \Omega_{C]}^\gamma \equiv - F_{ABC} + 3 \Omega_{[A,BC]}  \label{eq:GeneralisedCartanCurvatureExplicit} \\
	{\mathbf{\Theta}_{AB}}^\gamma &= - 2 \partial_{[A} \Omega_{B]}^\gamma - {f_{\alpha \beta}}^\gamma \Omega_A^\alpha \Omega_B^\beta - \tilde{\rho}^{ \alpha \gamma } \Theta_{\alpha AB} + {F^C}_{AB} \Omega_C^\gamma \nonumber \\
	{{\mathbf{\Theta}_A}^{\beta \gamma}} &= - \partial_A \rho^{\beta \gamma} - 2 \Omega^{B [{\beta} } \partial_B \Omega^{{\gamma}]}_A + \Omega^{B [{\beta}} \partial_A \Omega^{{\gamma}]}_B - {F_A}^{BC} \Omega_B^\beta \Omega_C^\gamma + 2 \Omega^{\delta }_A {f_{\delta \epsilon}}^{[\beta} \tilde{\rho}^{{\gamma}] \epsilon} \nonumber \\
	{\mathbf{\Theta}^{\alpha \beta \gamma}} &= 3 \Omega^{A [\alpha} \partial_A \rho^{\beta \gamma]} - 3 \Omega^{A[\underline{\alpha}} \Omega^{B \underline{\beta	}} \partial_A \Omega^{\underline{\gamma}]}_B + F^{ABC} \Omega_A^\alpha \Omega_B^\beta \Omega_C^\gamma + 3 \tilde{\rho}^{[\underline{\alpha}\delta} \tilde{\rho}^{\underline{\beta} \epsilon} f^{\underline{\gamma}]}{}_{\delta \epsilon}\,. \nonumber
\end{align}
They should be understood as curvatures or torsions, whereas the non-bold quantities correspond to the action of the symmetry. As in the ordinary Cartan connection case, ${\Theta_{\alpha B}}^C$ is defined as a component of the Cartan curvature but has no interpretation as curvature or torsion. Rather it defines the action of $H$ on the generalised tangent bundle in some generalised frame ${E_A}^M$. Nevertheless, it only depends on the model space algebra and the choice of a generalised frame ${E_A}^M$, not on the genuine connection part ($\Omega$ or $\rho$). Comparing our results with  \cite{Butter:2022iza}, one has to keep in mind that generalised Cartan geometry uses a non-holonomic basis for the generalised tangent space. As shown by \eqref{eq:actionOnCoordinates},  coordinates transform under $H$-gauge transformations and result in an action of the second index of the frame $E_B{}^L$ in \eqref{eq:nablaalphas}. In the holonomic basis used in \cite{Butter:2022iza} this does not happen. Here $H$ transformation only act on flat indices and do not mix with the generalised diffeomorphisms of the curved indices. Both conventions are connected by fixing $\nabla_\alpha {E_M}^A = 0$, or equivalently ${\Theta_{\alpha B}}^C \equiv {f_{\alpha B}}^C$. Doing so, we reproduce the formulae from \cite{Butter:2022iza} and the expected symmetry properties. Also, the relations \eqref{eq:JacobiIdentitiesH2} hold for ${\Theta_{\alpha \beta}}^\gamma$ and ${\Theta_{\alpha B}}^C$ as well.

\paragraph{Introducing non-vanishing components ${\Theta_{\alpha}}^{\beta \gamma}$ and ${\Theta_{\alpha B}}^\gamma$. } In the above derivation, these components vanish. From the world-sheet phase space point of view it seems logical to allow them, as they would correspond to terms proportional to $J_\alpha = s_\alpha \approx 0$. Indeed, these components can be introduced in the above derivation by assuming a non-standard, \textit{affine} transformation behaviour of $\Omega$ and $\rho$, namely
\begin{align}\label{eq:shiftOmega}
	\nabla_\alpha \Omega_B^\gamma &= - {\Theta_{\alpha B}}^C \Omega_C^\gamma + {f_{\alpha \beta }}^\gamma \Omega_B^\beta  \underline{- {f_{\alpha B}}^\gamma } \\ \label{eq:shiftRho}
	\nabla_\alpha \rho^{\beta\gamma} &= - 2 {f^{[\beta}}_{\alpha \delta} \rho^{\gamma] \delta } \underline{ - {f_{\alpha}}^{\beta \gamma} }.
\end{align}
If these formulae are used, we reproduce the results from \cite{Butter:2021dtu,Butter:2022iza}.

Moreover, non-vanishing ${\Theta_{\alpha B}}^\gamma$ or ${f_{\alpha B}}^\gamma$ means that the decomposition of the model algebra $\mathfrak{d} = \mathfrak{h} \oplus \tilde{\mathfrak{d}}$ is \textit{not reductive}, which is a typical assumption for Cartan geometry, as it ensured that the (generalised) Cartan connections identifies $\mathfrak{d}/(\mathfrak{h} + \mathfrak{h}^\star) \simeq (T \oplus T^\star)_{\pi(p)} M$.

\paragraph{Generalised Cartan Curvature for generic model algebras $\mathfrak{d}$.} For the most general case of a Lie algebra $\mathfrak{d}$ as model algebra with isotropic subalgebra $\mathfrak{h}$, all structure constants ${f^\mathcal{M}}_{KL}$ might be non-vanishing, apart from:
\begin{equation}
    {f_{\alpha\beta\gamma}} = 0 = {f_{\alpha \beta}}^C. 
\end{equation}
The form of the generalised Cartan connection \eqref{eq:GeneralisedCartanConnectionChoice} ensures that the same,
\begin{equation}
    {\Theta_{\alpha\beta\gamma}} = 0 = {\Theta_{\alpha \beta}}^C,
\end{equation}
holds also for the components of the Cartan curvature. Similar to the previous observation, we concern ourselves first with the components ${\Theta_{\alpha ...}}^{...}$. Structure constants $f$ will always the ones of the model Lie algebra  
\begin{align}
	{\Theta_{\alpha \beta}}^\gamma &= {f_{\alpha \beta}}^\gamma \nonumber, \qquad {\Theta_{\alpha B}}^C = {f_{\alpha B}}^C - {E_M}^C \nabla_\alpha {E_B}^M \nonumber \\
  {\Theta_{\alpha B}}^\gamma &= {f_{\alpha B}}^\gamma - \nabla_\alpha \Omega_B^\gamma - {\Theta_{\alpha B}}^C \Omega_C^\gamma + {f_{\alpha \beta}}^\gamma \Omega_B^\beta \label{eq:GeneralisedCartanCurvatureExplicitGeneric1} \\
	{\Theta_{\gamma}}^{\alpha \beta} &= {f_\gamma}^{\alpha \beta} - \nabla_\gamma \rho^{\alpha\beta} - 2 {f^{[\alpha}}_{\gamma\delta} \rho^{\beta] \delta} - 2 \Omega^{[\alpha}_A {\Theta_\gamma}^{\beta]A}  \label{eq:GeneralisedCartanCurvatureExplicitGeneric2}
\end{align}
Again, the components ${\Theta_{\alpha}}^{\beta \gamma}$ and ${\Theta_{\alpha B}}^\gamma$ could be absorbed into a non-trivial gauge transformation of $\Omega$ or $\rho$. If they transform as indicated in \eqref{eq:ConnectionGaugeRegular}, these components are given by
\begin{align}
    {\Theta_{\alpha B}}^\gamma &= {f_{\alpha B}}^\gamma, \qquad
	{\Theta_{\gamma}}^{\alpha \beta} = {f_\gamma}^{\alpha \beta} - 2 \Omega^{[\alpha}_A {f_\gamma}^{\beta]A}
\end{align}
after taking into account the model algebra structure constants and connection. For the genuine curvature and torsion components we furthermore find
\begin{align}
	{\mathbf{\Theta}}_{ABC} &= f_{ABC} - 3 {E_{[C}}^M \partial_A E_{B]M} + 3 \Theta_{\gamma [AB} \Omega_{C]}^\gamma  \nonumber \\
	{\mathbf{\Theta}_{AB}}^\gamma &=  - 2 \partial_{[A} \Omega_{B]}^\gamma - {f_{\alpha \beta}}^\gamma \Omega_A^\alpha \Omega_B^\beta - \tilde{\rho}^{ \alpha \gamma } \Theta_{\alpha AB} \nonumber \\
	&{} \qquad + {f_{AB}}^\gamma + ({F^C}_{AB} - {f^C}_{AB}) \Omega_C^\gamma + 2 \Omega^\alpha_{[A} {\Theta_{\alpha B]}}^\gamma \label{eq:GeneralisedCartanCurvatureExplicitGeneric3} \\
	{{\mathbf{\Theta}_A}^{\beta \gamma}} &= - \partial_A \rho^{\beta \gamma} - 2 \Omega^{B [{\beta} } \partial_B \Omega^{{\gamma}]}_A + \Omega^{B [{\beta}} \partial_A \Omega^{{\gamma}]}_B   + 2 \Omega^{\delta }_A {f_{\delta \epsilon}}^{[\beta} \tilde{\rho}^{{\gamma}] \epsilon} \nonumber \\
	&{} \qquad + ( {f_A}^{BC} - {F_A}^{BC} ) \Omega_B^\beta \Omega_C^\gamma + \Omega_C^\gamma {f_\gamma}^{\alpha \beta} + {f_C}^{\alpha \beta} + 2\Omega^{[\alpha}_B {f_C}^{\beta] B} + 2 {f_{\gamma}}^{C[\alpha} \tilde{\rho}^{\beta] \gamma} \\
	{\mathbf{\Theta}^{\alpha \beta \gamma}} &= 3 \Omega^{A [\alpha} \partial_A \rho^{\beta \gamma]} - 3 \Omega^{A[\underline{\alpha}} \Omega^{B \underline{\beta	}} \partial_A \Omega^{\underline{\gamma}]}_B + 3 \tilde{\rho}^{[\underline{\alpha}\delta} \tilde{\rho}^{\underline{\beta} \epsilon} f^{\underline{\gamma}]}{}_{\delta \epsilon}
	+ f^{\alpha \beta \gamma} \nonumber \\
	&{} \quad + (F^{ABC} - f^{ABC}) \Omega_A^\alpha \Omega_B^\beta \Omega_C^\gamma + 3 {\Theta_\delta}^{[\alpha \beta]} \tilde{\rho}^{\gamma]\delta} - 2 \Omega^{[\alpha}_B f^{\beta] BC} - 2 \Omega^{[\alpha}_A \tilde{\rho}^{\beta \delta} {f_\delta}^{\gamma A}
\end{align}
As demonstrated in \cite{Butter:2021dtu}, $\Theta^{\alpha \beta \gamma}$ does not consist of independent information -- it is completely fixed by the other components via Jacobi identities of the current algebra.

\paragraph{Dual approach.} One can arrive to the definition of generalised curvature \eqref{curvature} by more Cartan-inspired means. We start by writing a flat model Maurer-Cartan equation by dualising \eqref{zeromode}  
\begin{gather}
    \diff s^{\gamma} = {f_{\alpha \beta}}^\gamma s^{\alpha} \wedge s^{\beta} \nonumber
    \\
    \diff p^N = {f_{\alpha M}}^N s^{\alpha} \wedge p^M
    \\
    \diff \sigma_{\gamma} = {f_{ M N \gamma}} p^M \wedge p^N +  {{f_{\alpha}}^{\beta}}_\gamma s^{\alpha} \wedge \sigma_{\beta} \nonumber
\end{gather}
same can be done for more general algebras as discussed in the previous section but here we only study the generalised Poincar\'e case. When passing to the curved model we let the forms $\omega^{\mathcal{M}} = \{s,p,\sigma\}$ depend on the curved manifold coordinates according to 
 \eqref{eq:GeneralisedCartanConnectionChoice}.
 Explicitly we get
  \begin{equation} 
 \omega_{\text{curved}}^{\mathcal{M}} = \left( \begin{array}{ccc} \delta_\alpha^\mu & 0 & 0 \\ \Omega_A^\mu(X) & {E_A}^M(X) & 0 \\ \rho^{\alpha \mu}(X) - \frac{1}{2} \Omega^{B\alpha}(X) \Omega_B^\mu(X) & - \Omega^{\alpha}_A(X) {E^A}_M(X) & \delta^\alpha_\mu
	\end{array} \right) 
	\begin{pmatrix}
	s^{\alpha} \\ dx^A \\ \sigma_{\alpha}
	\end{pmatrix}
\end{equation}
The Maurer-Cartan equation for $\omega_{\text{curved}}$ now reads
\begin{gather}
    \diff \omega_{\text{curved}}^{\gamma} - {f_{\alpha \beta}}^\gamma \omega_{\text{curved}}^{\alpha} \wedge \omega_{\text{curved}}^{\beta} =\Theta^{\gamma} \, (=0) \nonumber
    \\
    \diff \omega_{\text{curved}}^N - {f_{\alpha M}}^N \omega_{\text{curved}}^{\alpha} \wedge \omega_{\text{curved}}^M = \Theta^N
    \\
    \diff \omega_{\text{curved}\, \gamma} - {f_{ M N \gamma}} \omega_{\text{curved}}^M \wedge \omega_{\text{curved}}^N +  {{f_{\alpha}}^{\beta}}_\gamma \omega_{\text{curved}}^{\alpha} \wedge \omega_{\text{curved}\, \beta} = \Theta_{\gamma} \nonumber
\end{gather}
where each of the components of the generalised Cartan curvature $(\Theta^{\gamma},\Theta^{N},\Theta_{\gamma})$ corresponds to \eqref{eq:GeneralisedCartanCurvatureExplicit}. \footnote{In the form language each of these $\Theta$ components is seen as a two form and \eqref{eq:GeneralisedCartanCurvatureExplicit} are the corresponding components when expanded in all indices explicitly. } And the vanishing of $\Theta^{\gamma}$ is in agreement with a special gauge of the curved Maurer-Cartan form.

\section{Generalised Cartan Geometry in Generalised Metric Formalism}\label{chap:genMetricFormalism}

So far, the Poláček-Siegel construction has only been employed in the generalised frame formalism \cite{Polacek:2013nla,Butter:2021dtu,Butter:2022iza}. Here, it will presented explicitly in the metric formalism. This is also done in order to clarify the geometric interpretation of the new quantities -- in particular, the new higher order tensors. The fact that the standard generalised torsion $\mathcal{T}_{ABC}$ and generalised Riemann curvature $\mathcal{R}_{KL,MN}$ appear as coefficients is remarkable. Consistency -- for the current algebra governed by the Jacobi identities -- imply that these and all the other are $H$-tensors.

\subsection{Generalised Cartan Curvature in terms of generalised Christoffel connection} \label{chap:GeneralisedMetric}

From now on, we will assume that
\begin{itemize}
    \item $H$ is the double Lorentz group O$(d) \times$O$(d)$ and
    \item the model algebra $\mathfrak{d}$ is the generalised Poincar\'e algebra whose current algebra realisation is given in \eqref{eq:CurrentAlgebraFlat}.
\end{itemize}
This is the setting suitable for connections $\Omega$ that are compatible with a generalised metric $\mathcal{H}_{MN}$ on the generalised tangent space. $H$ is then the reduced structure group on the generalised tangent bundle after fixing the generalised metric. The relevant components of the generalised Cartan connections are the ones computed in \eqref{eq:GeneralisedCartanCurvatureExplicit}.

There seem to be several ways to encode a generic Christoffel or affine connection in the generalised Cartan connection \eqref{eq:CartanConnectionChoice}. A direct definition of a Christoffel connection as ${\Gamma_{M,N}}^P = \Omega_M^\alpha {\Theta_{\alpha N}}^P$ and setting the generalised frame ${E_A}^M = \delta_A^M$ in the \eqref{eq:GeneralisedCartanConnectionChoice}, is possible and will yield the same results. Instead, we treat $\Omega$ as a spin connection and obtain the components of the generalised Cartan connection \eqref{eq:GeneralisedCartanCurvatureExplicit} in terms of a Christoffel connection via the vielbein postulate on the generalised vielbein, 
\begin{equation}
\nabla_I {E_A}^J = \partial_I {E_A}^J + {\Omega_{I,A}}^B {E_B}^J - {\Gamma_{I,K}}^J {E_A}^K = 0 \label{eq:VielbeinPostulate}\,
\end{equation}
form with we get
\begin{equation}
  \Gamma_{I,J}{}^K = E_J{}^A \partial_I E_A{}^K +  \Omega_{I,J}{}^K \,. 
\end{equation}
A similar algebraic relation can be introduced for the Poláček-Siegel field $\rho$. We denote the curved index (Christoffel) version of it as $\sigma$ with
\begin{align}
\sigma_{IJ,KL} &= \rho_{IJ,KL} + \frac{1}{2} \left( {E_I}^A (\partial_M E_{AJ})  {\Gamma^M}_{KL} - {E_K}^A (\partial_M E_{AL})  {\Gamma^M}_{IJ} \right) \nonumber \\
&= \rho_{IJ,KL} - \frac{1}{2} \left( \Omega_{M,IJ} {\Gamma^M}_{KL} - {\Omega_{M,KL}}  {\Gamma^M}_{IJ} \right) \label{eq:VielbeinPostulateRho}
\end{align}
where $\rho_{IJ,KL} = \rho^{\alpha \beta} {\Theta_{\alpha IJ}} {\Theta_{\beta KL}}$. In general, all objects $v^\alpha$ are written as $v_{KL} = v^\alpha \Theta_{\alpha KL}$ in the following. The components of generalised Cartan curvature \eqref{eq:GeneralisedCartanCurvatureExplicit} then take the following form
\begin{align}
\mathbf{\Theta}_{MNP} &= 3 \Gamma_{[M,NP]} \nonumber \\
\mathbf{\Theta}_{KL,MN} &= - r_{KL,MN} + \frac{1}{2} {\Gamma^Q}_{KL} \Gamma_{Q,MN} - \sigma_{KL,MN} \label{eq:GeneralisedCartanCurvatureGenMetric}\\
\mathbf{\Theta}_{P,KL,MN} &= - \nabla_P \sigma_{KL,MN} + \frac{1}{2}  \left( r_{QP,KL} {\Gamma^Q}_{MN} + (\partial_Q \Gamma_{P,KL}) {\Gamma^Q}_{MN} - (KL \leftrightarrow MN) \right). \nonumber
\end{align}
As mentioned above, $\mathbf{\Theta}_{KL,MN,PQ}$ is not an independent object \cite{Butter:2021dtu}. Hence, we will not introduce it here. $r_{KL,MN}$ is the ordinary curvature of $\Gamma$ given by
\begin{equation}
r_{KL,MN} = 2 \left( \partial_{[K} {\Gamma_{L],MN}} + {\Gamma_{[K,M}}^Q {\Gamma_{L],QN}} \right)
\end{equation}
This is the remarkable result of \cite{Polacek:2013nla} phrased in the generalised metric formalism:
\begin{itemize}
    \item $\mathbf{\Theta}_{MNP}$ is the so-called \textit{generalised torsion} $\mathcal{T}_{MNP}$ of a generalised Christoffel connection:
     \begin{equation}
         \mathbf{\Theta}_{MNP} \equiv \mathcal{T}_{MNP} = 3 \Gamma_{[M,NP]}
     \end{equation}
     We will simplify certain calculations below, by assuming that a torsion-free connection. 
    \item $\mathbf{\Theta}_{KL,MN}$ contains the \textit{generalised Riemann tensor} $\mathcal{R}_{KL,MN}$ \cite{Hohm:2011si} as its symmetric part:\footnote{In comparison to \cite{Hohm:2011si}, we use conventions in which $\Gamma \rightarrow - \Gamma$ and also, in addition, $r, \mathcal{R} \rightarrow -r , -\mathcal{R}$ }.
    \begin{align}
    \mathbf{\Theta}_{KL,MN} &= - \frac{1}{2} \left( \mathcal{R}_{KL,MN} + \mathcal{A}_{KL,MN} \right) \\
	\mathcal{R}_{KL,MN} &= - (\mathbf{\Theta}_{KL,MN} + \mathbf{\Theta}_{MN,KL}) = r_{KL,MN} + r_{MN,KL} - {\Gamma^Q}_{KL} \Gamma_{Q,MN}
	\end{align}
\end{itemize}
Here, we have shown the approach in \cite{Polacek:2013nla} can be understood as generalised Cartan geometry. Moreover, these objects, that are well-understood in the generalised geometry literature \cite{Jeon:2010rw,Coimbra:2011nw,Hohm:2011si}, appear as components of the generalised Cartan connection $\Theta$.

The key task in the remainder of this section is to understand the objects that appear as components of the generalised Cartan curvature but do not appear in the ordinary treatment of generalised geometry and double field theory. These are
\begin{itemize}
    \item the skewsymmetric part of $\mathbf{\Theta}_{KL,MN}$
\begin{align}
	\mathcal{A}_{KL,MN} &= - \mathbf{\Theta}_{KL,MN} + \mathbf{\Theta}_{MN,KL} = 2 \sigma_{KL,MN} + r_{KL,MN} - r_{MN,KL}\,,
\end{align}
    \item and the 5-index tensor $\mathbf{\Theta}_{P,KL,MN}$
\end{itemize}
in addition to the additional part $\sigma$ of the generalised Cartan connection, that does not have an equivalent in the standard treatment of generalised Riemannian geometry.

\paragraph{Anomalous behaviour under generalised diffeomorphisms.} The first step of investigation should be whether all objects are generalised tensors. Given our connection $\nabla_M V_N = \partial_M V_N + {\Gamma_{M,N}}^K V_K$, the anomalous transformation under generalised diffeomorphisms $\mathcal{L}_\xi$ of the Christoffel connection is
\begin{align}
    \Delta_\xi {\Gamma_{M,NK}} = - 2 \partial_{M} \partial_{[N} \xi_{K]}.
\end{align}
With this we can check explicitly that the components of the generalised Cartan curvature \eqref{eq:GeneralisedCartanCurvatureGenMetric} are indeed generalised tensors satisfying
\begin{equation}
    \Delta_\xi \mathbf{\Theta}_{KL,MN} = 0 , \qquad  \Delta_\xi \mathbf{\Theta}_{P,KL,MN} = 0
\end{equation}
part of which is the well-known $\Delta_\xi \mathcal{R}_{KL,MN} = 0$, with
\begin{equation}
    \Delta_\xi r_{KL,MN} = (\Delta_\xi {\Gamma^Q}_{MN} ){\Gamma}_{Q,KL}
\end{equation}
and implying that the so far unfixed transformation of $\sigma$ under generalised diffeomorphisms has to be fixed as
\begin{align}
  \Delta_\xi \sigma_{KL,MN} = - \frac{1}{2} ( \Delta_\xi r_{KL,MN} - \Delta_\xi r_{MN,KL} ).
\end{align}

\paragraph{Jacobi and Bianchi identities.} Certain consistency conditions between the components of $\Theta$ can be derived as Jacobi identities of the current algebra \eqref{eq:GeneralisedCartanCurvatureCurrent}. More details for parts of this discussion can be found in \cite{Butter:2022iza} for the generalised flux frame from the consistency conditions of extended double field theory.
\begin{itemize}
    \item The $(J_\alpha , J_M , J_N)$-Jacobi identity implies that $\mathbf{\Theta}_{KLM}$, $\mathbf{\Theta}_{KL,MN}$ and $\mathbf{\Theta}_{P,KL,MN}$ are $H$-tensors
    
    \item The $(J_K , J_L , J_M)$-Jacobi identity gives a generalisation of the \textit{Bianchi identities} for the generalised Riemann tensor:
    \begin{align}
        \mathcal{R}_{[KL,M]N} &= \text{generalised torsion terms} \\
        \nabla_{[P} \mathbf{\Theta}_{KL],MN} &= \mathbf{\Theta}_{[P,KL],MN} + \ \text{generalised torsion terms}
    \end{align}
    The second identity can be expressed, up to generalised torsion terms, as
    \begin{align}
        \frac{1}{2} \nabla_{[P} \mathcal{R}_{KL],MN} = - \left( \mathbf{\Theta}_{[P,KL],MN} + \frac{1}{2} \nabla_{[P} \mathcal{A}_{KL],MN} \right) \label{eq:DifferentialBianchi}
    \end{align}
    in terms of the generalised Riemann tensor. It is well-known \cite{Hohm:2011si} that, in general, there is no differential Bianchi identity of type $\frac{1}{2} \nabla_{[P} \mathcal{R}_{KL],MN} = 0$. Instead, the right-hand-side is a tensor that measures the violation of the differential Bianchi identity. This combination, that by construction defines a generalised tensor, will be denoted by
    \begin{equation}
         \Xi_{P,KL,MN} = \mathbf{\Theta}_{[P,KL],MN} + \frac{1}{2} \nabla_{[P} \mathcal{A}_{KL],MN}.
    \end{equation}
    As will be discussed in section \ref{chap:HigherTensors}, it is this combination that has clear geometric meaning. At this stage its relevance can be seen due to two reasons:
    \begin{itemize}
        \item $\Xi$ defines the differential Bianchi identity for the generalised Riemann tensor \eqref{eq:DifferentialBianchi}
        \item $\Xi$ is the unique combination of the novel tensors $\mathcal{A}_{KL,MN}$ and $\mathbf{\Theta}_{P,KL,MN}$ that does not depend on the Poláček-Siegel field $\sigma$. Hence, it is only defined in terms of generalised Christoffel connection and could contain genuine geometric information, although at higher (second) order in derivatives of the Christoffel connection.
    \end{itemize}
\end{itemize}
For a generic connection, there does not seem a relation between $\Xi$ and the other tensors we already understand. For physical applications concerned with the bosonic part of the supergravity action, one is however typically interested in generalised metric compatible connections. For them one can always introduce a torsion-less connection, called a \textit{generalised Levi-Civita connection}. In this case, we will be able to show that $\Xi$ is related to the generalised Riemann tensor and does not define new geometric information. To this end let us, briefly summarise the essentials of metric-compatible, torsion-free connections in generalised geometry.

\paragraph{Generalised Levi-Civita connection.} Generalising the Levi-Civita-connection in ordinary Riemannian geometry, to a generalised Levi-Civita connection is defined to be subject to three major constraints \cite{Hohm:2011si},
\begin{enumerate}
  \item compatibility with $\eta$: $\nabla_M \eta_{KL} = 0 \quad \Rightarrow \quad \Gamma_{M,NP} = - \Gamma_{M,PN}$\,,
  \item vanishing generalised torsion: $\mathcal{T}_{MNK} = 3 \Gamma_{[M,NK]} = 0$\,,
  \item Compatibility with a generalised metric $\mathcal{H}_{MN}$ and dilaton density.
\end{enumerate}
Still, this does not completely determine all components ${\Gamma_{M,N}}^P$ in terms of the generalized metric and dilaton. The trace-free part of $\Gamma_{\bar{M},\bar{N} \bar{P}}$ and $\Gamma_{\ubM,\ubN \ubP}$ is undetermined when introducing the standard decomposition with respect to the projectors $P_\pm = \frac{1}{2} \left( \eta \pm \mathcal{H} \right)$. Phrasing generalised Riemannian geometry in terms of generalised Cartan geometry will not resolve this issues of undetermined components. Instead, the connection decomposes into to a fixed part $\hat{\Gamma}$ and an undetermined part $\Sigma$
\begin{equation}
  \Gamma_{K,LM} = \hat{\Gamma}_{K,LM} + \Sigma_{\ubK,\ubL \ubM} + \Sigma_{\bar{K},\bar{L}\bar{M}}
\end{equation}
subject to 
\begin{equation}
	\Sigma_{ [\ubK, \ubL \ubM]} = 0 , \quad \text{and} \quad {\Sigma^{\ubK}}_{\ubL\ubK} = 0 
\end{equation}
and the same for $\Sigma_{\bar{K},\bar{L}\bar{M}}$. Besides that, two important  consequences of metric compatibility are
\begin{align}
  r_{KL,\ubM\bN} &\equiv 0, \qquad \text{and} \qquad 
  \Gamma_{L,\ubM\bN} = (\bP \partial_L P)_{\bN \ubM} \label{eq:MetricComp}
\end{align}
The latter property will be particularly relevant for the contraction ${\Gamma^L}_{\ubM\bN} \partial_L f = 0$ acting on a general function $f$ due to the section condition.

\subsection{Higher Tensors for the Generalised Levi-Civita Connection} \label{chap:HigherTensors}
Generalised Cartan geometry, as introduced above, contains three objects that deserve further clarification:
\begin{itemize}
    \item The additional part $\rho$, or $\sigma$ respectively, of the Cartan connection that has no obvious geometric interpretation.
    \item The component $\mathcal{A}_{KL,MN} = \mathbf{\Theta}_{MN,KL} - \mathbf{\Theta}_{KL,MN}$ of the Cartan curvature.
    \item The component $\mathbf{\Theta}_{P,KL,MN}$ of the Cartan curvature. 
\end{itemize}

\paragraph{The $\mathcal{A}$-tensor as torsion for $\rho$.} Our analysis here will depend on $H$ being the double Lorentz group O$(d)\times$O$(d)$, mainly due to its dimension $2\times \frac{d(d-1)}{2}$. Rewriting $\rho_{IJ,KL} = \rho^{\alpha \beta} \Theta_{\alpha IJ} \Theta_{\beta KL}$ in terms of generalised tangent space indices obfuscates counting of degrees of freedom of $\rho$. In fact, it only has
\begin{equation}
    \# \rho_{IJ,KL} = \# \rho^{\alpha \beta} = \frac{1}{2}  (d^2 -d ) (d^2 - d  - 1) = \frac{1}{2} ( d^4 - 2d^3 + d) . \label{eq:dofrho}
\end{equation}
degrees of freedom. A similar analysis for $\sigma$ is not as clear but luckily it will not be needed apart from the relation \eqref{eq:VielbeinPostulateRho}. In particular, if $\nabla_\alpha E = 0$, then 
\begin{align}
    \Theta_{\alpha IJ} = f_{\alpha IJ} = f_{\alpha \ubI \ubJ} + f_{\alpha \bI \bJ}
\end{align}
from \eqref{eq:ConnectionGaugeRegular}. The last identity is the definition of the action of O$(d) \times $O$(d)$ on the generalised tangent space, as \cite{Polacek:2013nla} and results in the independent components
\begin{equation}
    \rho_{\bI\bJ,\bK\bL}, \quad \rho_{\bI\bJ,\ubK\ubL}, \quad \rho_{\ubI \ubJ,\ubK \ubL}
\end{equation}
of $\rho$, whereas $\rho_{\bI\ubJ,KL} = \rho_{KL,\bI\ubJ} = 0$ coinciding with \eqref{eq:dofrho}. $\mathcal{A}$ on the other hand is, a priori, a generic tensor with its index structure. Hence, its number of independent components is
\begin{equation}
    \# \mathcal{A}_{IJ,KL} = \frac{1}{2} \left( \frac{2d(2d-1)}{2} \left( \frac{2d(2d-1)}{2} - 1 \right) \right) = \frac{1}{2} ( 4d^4 - 4d^3 - d^2 + d) > \# \rho_{IJ,KL}. \nonumber
\end{equation}
Consequently, one only can fix some components of $\mathcal{A}$ to vanish by choosing $\rho$ appropriately, but not all. More precisely, what can be fixed are the components
\begin{equation}
    \mathcal{A}_{\bI\bJ,\bK\bL}=0, \quad \mathcal{A}_{\bI\bJ,\ubK\ubL} = 0, \quad \mathcal{A}_{\bI\bJ,\ubK\ubL} = 0
\end{equation}
after imposiong
\begin{equation}
    \sigma_{\bI\bJ,\bK\bL} = r_{\bI\bJ,\bK\bL} - r_{\bK\bL,\bI\bJ}.
\end{equation}
In this sense, the field $\mathcal{A}$ could be considered as the torsion for $\sigma$ and thereby $\rho$. This is the fixing of $\rho$, already performed in the generalised frame formalism in \cite{Polacek:2013nla}. What can be additionally understood in a new way here is what the remaining components of $\mathcal{A}$ are. First, one can quickly note that
\begin{align}
\mathcal{A}_{\ubI \bJ , \ubK \bL} \equiv 0
\end{align}
due to \eqref{eq:MetricComp}. Then
\begin{align}
\mathcal{A}_{\bI \ubJ , \ubK \ubL} &= - r_{\bI \ubJ, \ubK \ubK} + 2 \sigma_{\bI \ubJ, \ubK \ubK} = - \mathcal{R}_{\bI \ubJ, \ubK \ubK} - {\Gamma^Q}_{\bI \ubJ} \Gamma_{Q,\ubK \ubL} + ({E_{\bI}}^A {P_L}^P \partial_Q E_{PA}) {\Gamma^Q}_{\ubK \ubL} \nonumber\\
&= \mathcal{R}_{\bI \ubJ, \ubK \ubK} - {\Gamma^Q}_{\bI \ubJ} \Gamma_{Q,\ubK \ubL} + (- \Omega_{Q,\ubK \ubL} + \Gamma_{Q,\ubK \ubL}){\Gamma^Q}_{\ubK \ubL} =\mathcal{R}_{\bI \ubJ , \ubK \ubL}
\end{align}
using the vielbein postulate \eqref{eq:VielbeinPostulate} and the section condition multiple times. Similarly,
\begin{equation}
\mathcal{A}_{\ubI \bJ , \bK \bL} = \mathcal{R}_{\ubI\bJ , \bK \bL}\,.
\end{equation}
Hence, while $\mathcal{A}_{\ubI \bJ , \bK \ubL}$ vanishes automatically, $\mathcal{A}_{\ubI \bJ , \bK \bL}$ and $\mathcal{A}_{\bI \ubJ , \ubK \ubL}$ cannot be set to zero without putting a constraint on the generalised Riemann curvature. We conclude that neither $\sigma$ nor $\mathcal{A}$ contain new geometric information. Most of the latter's components can be given arbitrary values by an appropriate choice of $\sigma$ because $\mathcal{A}$ contains the torsion of $\sigma$. The remaining components corresponding to certain projections of the generalised Riemann tensor.

\paragraph{$\mathcal{A}$ and $\sigma$ for arbitrary gauge groups $H$.} For a generic gauge group this decomposition of $\Theta$ not as insightful. Still a little bit can be done when the action of $H$ on the generalised tangent space is faithful. In this case, the exists a tensor $\widehat{f}^{\alpha B C}$ with the defining property
\begin{equation}
    \widehat{f}^{\alpha CD} f_{\beta CD} = \delta_\beta^\alpha\,.
\end{equation}
It allows to decompose at least parts of $\mathbf{\Theta}^{(4)} \sim {\mathbf{\Theta}^{\alpha}}_{AB}$ as
\begin{equation}
    \mathbf{\Theta}^{(4)} \sim \mathbf{\Theta}^{\alpha,\beta} + ... = \mathcal{A}^{\alpha\beta} + \mathcal{S}^{\alpha\beta} + ..., \qquad \text{with} \qquad \mathbf{\Theta}^{\alpha,\beta} = {\mathbf{\Theta}^\alpha}_{CD} t^{\beta CD},
\end{equation}
showing that parts of it will be $\mathfrak{h}$-representation. Moreover, there is always a bivector-like part $\mathcal{A}^{\alpha \beta}$ that can be fixed by a choice of $\rho^{\alpha \beta}$, or vice versa. Hence, $\rho$ and part the correspond part of $\mathbf{\Theta}^{(4)}$ do not contain geometric information. But as shown by the check of covariance with respect to generalised diffeomorphisms, the inclusion of $\rho$ is necessary to render the full curvature component $\mathbf{\Theta}_{IJ,KL}$ in  \eqref{eq:GeneralisedCartanCurvatureGenMetric} covariant.

\paragraph{The $\Xi$-tensor and its relation to the generalised Riemann tensor.} The auxiliary part $\sigma$ drops out of the combination $\Xi =  \mathbf{\Theta}^{(5)} + \frac{1}{2} \nabla \mathcal{A}$, as discussed above, such that we are left with
\begin{equation}
    \Xi_{P,KL,MN} = \frac{1}{2} \left( \nabla_{P} r_{MN,KL} + {\Gamma^Q}_{MN} \left(r_{QP,KL} + \partial_Q \Gamma_{P,KL} \right) \right) - (KL \leftrightarrow MN).
\end{equation}
When fixing $\mathcal{A}$ as above, one can basically identify the discussions about $\Xi$ and $\mathbf{\Theta}^{(5)}$ because components of both tensors will only differ by covariant derivatives of the generalised Riemann tensor.

For a generic connection $\Xi$ seems to be a new covariant object that can be built out of the connection. For a generalised Levi-Civita connection, on the other hand, in particular with its properties \eqref{eq:MetricComp}, it turns out that almost all components (projections) of $\Xi$ are determined in terms of the generalised Riemann tensor, namely
\begin{align}
    \Xi_{P,\ubK\bL,\ubM\bN} &= 0, \\
   \Xi_{P,\ubK\ubL,\ubM\bN} &= \frac{1}{2} \nabla_P \mathcal{R}_{\ubK\ubL,\ubM\bN}, \label{eq:XiIdentities}\\
   \qquad \Xi_{\ubP,\bK\bL,\ubM\ubN} &= - \frac{1}{2} \nabla_{\ubP} \mathcal{R}_{\bK\bL,\ubM\ubN} \\
    \Xi_{\bP,\ubK\ubL,\ubM\ubN} &= + \frac{1}{2} \left( \nabla_{[\ubK} \mathcal{R}_{\ubL]\bP,\ubM\ubN} - \nabla_{[\ubM} \mathcal{R}_{\ubN]\bP,\ubK\ubL} \right)\,, \label{eq:XiIdentities2}
\end{align}
and similar relations for components with the exchange of overbarred and underbarred indices. These calculations are significantly simplified when employing the differential Bianchi identity \eqref{eq:DifferentialBianchi}. As an example for the necessary calculations that involve the intrinsic proporties of a generalised Levi-Civita connection let us present the derivation of \eqref{eq:XiIdentities}:
\begin{align*}
   &{} \qquad \Xi_{P,\ubK\ubL,\ubM\bN} \\
     &= \frac{1}{2} \left( \nabla_P r_{\ubM\bN,\ubK\ubL} + r_{QP,\ubK\ubL} {\Gamma^Q}_{\ubM\bN} + \cancel{{\Gamma^Q}_{\ubM\bN} \partial_Q \Gamma_{P,\ubK\ubL}} - \cancel{\nabla_P r_{\ubK\ubL,\ubM\bN}} - \cancel{r_{QP,\ubM\bN} {\Gamma^Q}_{\ubK\ubL}} - {\Gamma^Q}_{\ubK\ubL} \partial_Q \Gamma_{P,\ubM\bN} \right) \\
    &= \frac{1}{2} \left( \nabla_P (\mathcal{R}_{\ubK\ubL,\ubM\bN} + {\Gamma^Q}_{\ubK\ubL} {\Gamma_{Q,\ubM\bN}})  + r_{QP,\ubK\ubL} {\Gamma^Q}_{\ubM\bN} - {\Gamma^Q}_{\ubK\ubL} \partial_Q \Gamma_{P,\ubM\bN} \right) \\
    &= \frac{1}{2} \left( \nabla_P \mathcal{R}_{\ubK\ubL,\ubM\bN} - \cancel{ {\Gamma^Q}_{\ubK \ubL} r_{QP,\ubM \bN}} \right)\,.
\end{align*}
Note that the third line can be obtained from the second one by writing $r$ and $\partial \Gamma$ explicitly in terms of $\nabla \Gamma$- and $\Gamma \Gamma$-terms.

In general, one observes that all these components are related to covariant derivatives of the generalised Riemann tensor. Hence, they do not describe independent geometric information. The only projections that are not captured by the above list are $\Xi_{\ubP,\ubK\ubL,\ubM\ubN}$ and $\Xi_{\bP,\bK\bL,\bM\bN}$. Indeed, it turns out that these cannot be written as a covariant derivative of $\mathcal{R}$. The most general ansatz that would be compatible with the index structure of these components is of the form \eqref{eq:XiIdentities2}, and thereby proportional to $\nabla_{[\ubK} \mathcal{R}_{\ubL]\ubP,\ubM\ubN} - \nabla_{[\ubM} \mathcal{R}_{\ubN]\ubP,\ubK\ubL}$. One quickly checks from  that the $\partial \partial \Gamma$ terms of $\Xi_{\ubP,\ubK\ubL,\ubM\ubN}$ take the form
\begin{align}
    \Xi_{\ubP,\ubK\ubL,\ubM\ubN} \Big\vert_{\partial \partial \Gamma} = -  \partial_{\ubP} \partial_{[\ubK} \Gamma_{\ubL],\ubM \ubN} + \partial_{\ubP} \partial_{[\ubM} \Gamma_{\ubN],\ubK \ubL}
\end{align}
which does not agree with this ansatz. So, $\Xi_{\ubP,\ubK\ubL,\ubM\ubN}$ and $\Xi_{\bP,\bK\bL,\bM\bN}$ seem to define a new tensor \textit{independent of the generalised Riemann tensor} (at least not proportional to its covariant derivatives). On the other hand, already the form of the $\partial\partial \Gamma$-terms makes it clear that these components depend on the unphysical components $\Sigma_{\ubL,\ubM\ubN}$ of $\Gamma$.

\paragraph{Physical and unphysical parts of $\Xi$.} The above relations imply in particular
\begin{equation}
    {\Xi^{\bN}}_{[\ubK\ubL,\ubM]\bN}=0
\end{equation}
by the algebraic Bianchi identity for $\mathcal{R}$ and additionally for the trace
\begin{equation}
    {\Xi^{\bN}}_{\ubK\ubL,\ubM\bN} \eta^{\ubL\ubM} = - \frac{1}{2} \nabla^{\bP} \mathcal{R}_{\ubK \bP}
\end{equation}
with the generalised Ricci tensor $\mathcal{R}_{\ubM \bN} = {\mathcal{R}_{\bL\ubM , \bN}}^{\bL}$. Hence, the traceless part of this component
\begin{equation}
    X_{\ubK \ubL,\ubM} = {\Xi^{\bN}}_{\ubK\ubL,\ubM\bN} + \frac{1}{D-1} {\nabla^{\bN}} \mathcal{R}_{\bN [ \ubK} P_{\ubL] \ubM}
\end{equation}
has exactly the index structure of the undetermined part of the connection. Explicitly, one can compute that
\begin{equation}
    X_{\ubK \ubL,\ubM} = -\frac{1}{2} \left( \nabla^{\bN} \nabla_{\bN} \Sigma_{\ubM,\ubK \ubL} + \hat{\nabla}^{\bN} \hat{\mathcal{R}}_{\bN \ubM,\ubK \ubL} + \frac{1}{D-1} \hat{\nabla}^{\bN} \hat{\mathcal{R}}_{\bN [ \ubK} P_{\ubL] \ubM}\right)
\end{equation}
where hatted quantities $\hat{\mathcal{R}}$, $\hat{\Gamma}$ are the ones defined in terms of the determined part of the connection. This demonstrates that $\Xi$ contains dependence on both physical and unphysical (undetermined by generalised metric or dilaton) parts of the connection.

On a more speculative note, putting $X = 0$ will partially fix the undetermined component without putting constraints on the geometry or breaking generalised covariance. Therefore, one is tempted to identify
\begin{equation}
     X_{\ubK \ubL,\ubM}, X_{\bK \bL,\bM} = 0, \quad \leftrightarrow \quad \text{Equation of motion for $\Sigma_{\ubM,\ubK \ubL}, \Sigma_{\bM, \bK \bL}$.}
\end{equation}
It would give (second order) dynamics of the undetermined components in terms of the geometrically determined components of the generalised Riemann and Ricci tensor. For each component $X$ there is one component of $\Sigma$ that could be fixed to make it vanish. In particular, for flat space (in the determined parts of the connection) one would have the freedom to add harmonic functions in the undetermined components. Equations of motion like that might be useful to constrain possible field redefinitions that were conjectured to be necessary to eliminate the dependence of higher derivative terms on the undetermined part of the connection \cite{Hohm:2011si}.

\section{Covariant String Dynamics} \label{chap:CovariantStringDynamics}
Generalised Cartan geometry is naturally realised on the phase space of a gauged 2d $\sigma$-model. In particular, it is the natural framework to describe dynamics both covariantly under
\begin{itemize}
    \item generalised diffeomorphisms, and
    \item an additional gauge symmetry with a gauge group $H$ that acts on the phase space as \eqref{eq:GeneralisedCartanGeometryAction}.
\end{itemize}
The construction in this section is a generalisation of the one in \cite{Lacroix:2023ybi}, where the dynamics of a gauged $\mathcal{E}$-models was considered. In the language of generalised Cartan geometry, these models describe $\sigma$-models with target spaces that are dressing cosets \cite{Klimcik:1996np} $G \backslash D / H$ for some doubled Lie group $D$ that corresponds to our model quasi-Lie bialgebra $\mathfrak{d}$. Generalised Cartan geometry allows to generalise this to arbitrary target spaces, by introducing a generalised Cartan connection. The motivating example still is generalised Cartan geometry with the generalised Poincar\'e algebra as model algebra, but the procedure could be easily extended to arbitrary model algebras $\mathfrak{d}$.

Let us consider the phase space described by the currents $J_\mathcal{A} (\sigma)$ with the current algebra\footnote{Just as reminder for the reader: This is the Poisson bracket on the phase space, but up to world-sheet boundary terms it also describes a Dorfman bracket.}
\begin{align*}
    \{ J_\mathcal{A}(\sigma_1) , J_\mathcal{B}(\sigma_2) \} = \eta_{\mathcal{AB}} \delta^\prime(\sigma_1 - \sigma_2) + {\Theta^\mathcal{C}}_{\mathcal{AB}}(\sigma_1) J_\mathcal{C}\delta(\sigma_1 - \sigma_2)
\end{align*}
given in \eqref{eq:GeneralisedCartanCurvatureCurrent}. Moreover, as generators of a gauge symmetry, the $J_\alpha$ are constraints and are set to zero,
\begin{equation}
	J_\alpha \approx 0.
\end{equation}
In the following, we will analyse the Hamiltonian dynamics of this system.

\subsection{Characterisation of a Hamiltonian}
Similarly to \cite{Lacroix:2023ybi}, we will identify $\mathfrak{h}^\star \sim \text{span}(J_\alpha)$ such that the currents decompose in the same way as the generators of the model algebra $\mathfrak{d}$. This implies
\begin{equation}
	\text{current phase space variables} \in \mathfrak{h} \oplus \mathfrak{h}^\star \oplus V,
\end{equation}
where $V$ is spanned by $J_A$. The setting here is slightly more general, in the sense that in the current algebra \eqref{eq:GeneralisedCartanCurvatureCurrent} the components of $\Theta$ do not need to be constant as it was required in  \cite{Lacroix:2023ybi} to obtain gauged $\mathcal{E}$-models. Nevertheless, in its core the logic of \cite{Lacroix:2023ybi} applies here, too. Natural assumptions for a Hamiltonian are:
\begin{itemize}
	\item It should be quadratic in currents. Hence, it would be characterised by a symmetric bilinear form $\mathcal{H}^{\mathcal{A}\mathcal{B}}$ (the generalised metric). In order to connect to the $\mathcal{E}$-model literature one can consider the operator form $(\hat{\mathcal{H}})^\mathcal{A}{}_\mathcal{B}$, where the last index is lowered with the O$(d',d')$-metric, as the $\mathcal{E}$-operator.
	
	\item The Hamiltonian vanishes on the auxiliary currents $J^\alpha$ which leads to the constrain \cite{Lacroix:2023ybi}
		\begin{align*}
			\text{Ker}(\hat{\mathcal{H}})\vert_{\mathfrak{h} \oplus V} = \mathfrak{h}\,.
		\end{align*}
Combining it with $J_\alpha \approx 0$ we find that the Hamiltonian can be expressed as
\begin{equation}
	H = \frac{1}{2} \int \mathrm{d} \sigma \ \mathcal{H}^{AB} J_A J_B\,,
\end{equation}
and therefore only includes the internal $O(d,d)$-currents. This is logical, as there should not be more information in the Hamiltonian when forgetting about the gauge symmetry. Our parameterisation \eqref{eq:GeneralisedCartanConnectionChoice} of generalised Cartan connection is exactly such that the Hamiltonian defined by these properties keeps that form \textit{weakly}:
\begin{align}
H &= \frac{1}{2} \int \mathrm{d}\sigma \ \mathcal{H}^{MN}(x) P_M (\sigma) P_N(\sigma) \nonumber \\
&= \frac{1}{2} \int \mathrm{d}\sigma \ \mathcal{H}^{AB} (J_A (\sigma)+ \Omega_A^\alpha(x) J_\alpha (\sigma) ) ( J_B (\sigma)+ \Omega_B^\beta J_\beta (\sigma)) \nonumber \\ 
&\approx \frac{1}{2} \int \mathrm{d} \sigma \ \mathcal{H}^{AB} J_A (\sigma) J_B(\sigma) \nonumber
\end{align}		
		
	\item The Hamiltonian should respect the gauge symmetry $H$, implying
	\begin{equation}
		\{J_\alpha , H \} = 0\,.
	\end{equation}
	Explicitly, in the generalised flux frame this is
	\begin{equation}\label{eq:mathcalHinv}
		\mathcal{H}^{A(B} {\Theta_{\alpha A}}^{C)} = 0
	\end{equation}
	and phrased in \cite{Lacroix:2023ybi} as: $\hat{\mathcal{H}}$ commutes with the adjoint action of $H$.
	
	\item $\mathcal{H}$ is an element of O$(d,d)$. In operator notation this translates to $\mathcal{\hat{H}}^3 = \mathcal{\hat{H}}$, which is the degenerate version of $\mathcal{\hat{H}}^2 = 1$ for the non-generate $\mathcal{E}$-model.
	
		\item 
	All parts of the Hamiltonian coupling to the currents $J_\alpha$ generate $H$-gauge transformations and are thus not relevant, leading to ${\mathcal{H}_\mathcal{A}}{}^\beta \approx 0$. But just removing them from the Hamiltonian would result in a loss of manifest gauge-covariance for the equations of motion. Therefore, instead of parameterise all different but redundant Hamiltonian, we restore manifest gauge-covariance by adding a Lagrangian multiplier $\mu^\alpha$ to the Hamiltonian, resulting in
	\begin{equation}
		H = \int \mathrm{d}\sigma \left( \frac{1}{2}	J_A J_B \mathcal{H}^{AB} + \mu^\alpha J_\alpha \right). \label{eq:HamiltonianMultiplier}
\end{equation}
One of the key observations of\cite{Lacroix:2023ybi} in our context is that the auxiliary current $J^\alpha$ and the Lagrange multiplier (or momentum map) field $\mu^\alpha$ are components of a $\mathfrak{h}$-valued gauge field. This will become clearer in the study of the covariant equations of motion.
\end{itemize}

\subsection{The equations of motion}
As equations of motion $\partial_\tau J_\mathcal{A} = \{ J_\mathcal{A} , H \}$ for the Hamiltonian \eqref{eq:HamiltonianMultiplier} one finds
\begin{align*}
\partial_\tau J_\alpha &\approx 0, \qquad\qquad \partial_\tau \mu^\alpha \approx 0 \\
\partial_\tau J_A &= \eta_{AB} \mathcal{H}^{BC} \partial_\sigma J_C - \mu^\alpha {\Theta_{\alpha A}}^B J_B + \Theta_{\beta AC} J^\beta J_C \mathcal{H}^{BC} + \mathbf{\Theta}^D{}_{AC} J_D \\
\partial_\tau J^\alpha &\approx \partial_\sigma \mu^\alpha + {f^\alpha}_{\beta \gamma} J^\beta \mu^\gamma + {\mathbf{\Theta}^\alpha}_{BC} J^C J_D \mathcal{H}^{BD}
\end{align*}
These equations correspond to a, in general, non-flat $H$-connection,whose curvature is measured by the curvature of the Cartan connection as
\begin{equation}
	\mathrm{d} A^\alpha + [A , A]^\alpha = {\mathbf{\Theta}^\alpha}_{BC} \mathbf{J}^B \wedge \mathbf{J}^C
\end{equation}
Here, we use the components of this new $H$-connections
\begin{equation}
    A^\alpha = (A_0^\alpha , A_1^\alpha) = (\mu^\alpha , J^\alpha)\,,
\end{equation}
and the components of the current one-form
\begin{equation}
    \mathbf{J}^A_0 = \mathcal{H}^{AB} J_B, 
    \qquad \text{and} \qquad 
    \mathbf{J}^A_1 = J^A\,,
\end{equation}
which follow from the typical self-duality relation $\mathbf{J}_C = \star \mathcal{H}_{CD} \mathbf{J}^D$ of duality covariant string dynamics. The covariant derivative corresponding to $A^\alpha$ is defined as
\begin{equation}
    \nabla = d + [A, \cdot]\,.
\end{equation}
We rewrite the equations of motion in the manifestly covariant form
\begin{equation}\label{eqn:masterCurrentAlgebra}
    \boxed{
    \begin{aligned}
        \mathrm{d} A^\alpha + [A , A]^\alpha &= {\mathbf{\Theta}^\alpha}_{BC} \mathbf{J}^B \wedge \mathbf{J}^C \\
        \nabla \mathbf{J}^A + \mathbf{\Theta}^A{}_{BC} \mathbf{J}^B \wedge     \mathbf{J}^C &= 0    
    \end{aligned}
    }
\end{equation}
This gives the component ${\mathbf{\Theta}^\alpha}_{BC}$ and $\mathbf{\Theta}_{ABC}$ of the generalised Cartan connection a clear interpretation for string dynamics. Its other parts do not appear, at least at the classical level. Note that this is a reformulation of the standard Hamiltonian approach to the bosonic string, or any 2d $\sigma$-model for that matter. It does not require any additional restrictions. 

From a geometric point of view, this simply a generalised geodesic equation, phrased in terms of general (torisonful) connection. The first equation in \eqref{eqn:masterCurrentAlgebra} is the definition of the curvature 

A distinguishing feature of this rewriting is its duality invariance. All known generalised dualities will leave the Hamiltonian \eqref{eq:HamiltonianMultiplier} invariant, whereas the current algebra \eqref{eqn:masterCurrentAlgebra} transforms covariantly \cite{Sfetsos:1997pi,Osten:2019ayq,Demulder:2019vvh,Borsato:2021gma,Butter:2022iza}. This encodes that the dual target spaces results in the same dynamics is completely manifest while on the level of the metric and the $B$-field complicated transformations are required.  

\paragraph{The irrelevance of $\rho$ for the classical dynamics.}
We see that the tensor ${\mathbf{\Theta}_C}^{\alpha \beta}$ drops out of the equations of motion. Additionally, the same happens for $\rho$ in ${\mathbf{\Theta}^\gamma}_{AB}$ because of $\partial_\tau J_\alpha = 0$. It implies 
\begin{equation}
	{\mathbf{\Theta}^\alpha}_{BC} J^C J_D \mathcal{H}^{BD} = ... + \rho^{\alpha \beta} {\Theta_{\beta B}}^{(C} \mathcal{H}^{D)B} J_C J_D = ... + 0
\end{equation}
after taking into account \eqref{eq:mathcalHinv}.

\paragraph{Chiral current algebra.} When ${\mathbf{\Theta}^\alpha}_{BC} J^B \mathcal{H}^{CD} J_D = 0$ holds, the connection $A$ is flat and we can easily integrate it out, for example by $A\approx 0$ and therefore $\mu^\alpha \approx 0 \approx J^\alpha$. In this is case we impose the vanishing of the connection $A$ as a constraint and need to employ the Dirac procedure to obtain the brackets for the remaining currents. But let us note, that this a very special case where there is no coupling to the background curvature at all. It drops out of the equations of motion and contains the chiral case considered in \cite{Lacroix:2023ybi}.

Imposing the constraints
\begin{equation}
	J_\alpha \approx 0, \qquad J^\alpha \approx 0\,,
\end{equation}
requires the Dirac procedure because the constraint algebra does not close weakly. Combining the two indices on both currents into ${}_a = ({}_\alpha , {}^\alpha)$, we obtain the brackets
\begin{equation}
	\{J_a (\sigma_1) , J_b (\sigma_2) \} \approx \kappa_{ab}(\sigma_1,\sigma_2)
\end{equation}
with
\begin{equation}
	\kappa_{ab}(\sigma_1,\sigma_2) = \left( \begin{array}{cc} 0 & \delta_\alpha^\beta \delta^\prime(\sigma_1 - \sigma_2) \\
	\delta_\beta^\alpha \delta^\prime(\sigma_1 - \sigma_2) & {\mathbf{\Theta}^{\alpha \beta}}_{C} J^C (\sigma_1) \delta(\sigma_1 - \sigma_2) \end{array} \right).
\end{equation}
The inverse of this matrix is
\begin{equation}
	(\kappa^{-1})_{ab}(\sigma_1 , \sigma_2) = \left( \begin{array}{cc} 0 & \delta_{\alpha}^\beta \epsilon(\sigma_1 - \sigma_2) \\
	- \delta^\alpha_\beta \epsilon(\sigma_2 - \sigma_1) & \tilde{J}^{\alpha \beta} (\sigma_1 , \sigma_2)
	\end{array} \right)
\end{equation}
with the non-local contribution\footnote{Using string world-sheet conventions $0 \leq \sigma \leq 2\pi$.}
\begin{equation}
	\tilde{J}^{\alpha\beta}(\sigma_1, \sigma_2) = \int^{\text{min}(\sigma_1,\sigma_2)}_0 {\mathbf{F}^{\alpha \beta}}_{C} J^C(\sigma) \ \mathrm{d}\sigma.
\end{equation}
Hence, we find the non-vanishing Dirac bracket
\begin{equation}\label{eq:DiracBracket}
    \begin{aligned}
	    \{ J_A (\sigma_1) , J_B (\sigma_2) \}_{D.B.} &\approx \eta_{AB} \delta^\prime(\sigma_1 - \sigma_2) + \mathbf{\Theta}_{ABC} J^C(\sigma_1) \delta(\sigma_1-\sigma_2) \\
	    &{} \quad + \Big( {\mathbf{\Theta}^\gamma}_{BD} \Theta_{\gamma AC} \ \epsilon(\sigma_1-\sigma_2) - {\mathbf{\Theta}^\gamma}_{AC} \Theta_{\gamma BD} \ \epsilon(\sigma_2-\sigma_1) \Big) J^C(\sigma_1) J^D(\sigma_2) \\
	    &{} \quad + {\mathbf{\Theta}^{\gamma\delta}}_E \Theta_{\gamma AC} \Theta_{\delta BD} J^C(\sigma_1) J^D(\sigma_2) \tilde{J}^E(\sigma_1,\sigma_2)
    \end{aligned}
\end{equation}
Remarkably, one finds a dependence on the higher tensor $\mathbf{\Theta}_{P,KL,MN}$ here. Assuming that these components of the generalised Cartan curvature are constant, one obtains as projection to the zero modes
\begin{align*}
	\{ J_A , J_B \} &= {\mathbf{\Theta}}_{ABC} J^C + ({\mathbf{\Theta}^\gamma}_{BD} f_{\gamma AC} + {\mathbf{\Theta}^\gamma}_{BC} f_{\gamma AD}) \tilde{J}^{CD} + \mathbf{\Theta}_{E,AC,BD} \tilde{J}^{CD,E} 
\end{align*}
with the non-local contributions
\begin{equation}
	\tilde{J}^{AB} = \int_0^{2\pi} \mathrm{d}\sigma_1 \  J^A(\sigma_1) \int^{\sigma_1}_0 \mathrm{d}\sigma_2 \ J^B(\sigma_2), \qquad \tilde{J}^{AB,C} = \int_0^{2\pi} \mathrm{d} \sigma_1 \int_0^{2\pi} \mathrm{d} \sigma_2 \ J^A(\sigma_1) J^B(\sigma_2) \tilde{J}^C (\sigma_1,\sigma_2)\,.
\end{equation}
Nevertheless, in the equations of motion the contributions of $\mathbf{\Theta}_{P,KL,MN}$ drops out again.

\section{Outlook}\label{chap:Outlook}
In this article, we introduced generalised Cartan geometry and demonstrated how it permits a fully duality covariant formulation of classical 2d $\sigma$-model dynamics. An important advantage compared to Riemannian generalised geometry is that covariant curvature tensors arise automatically. However, the problem of undetermined of the connection which cannot be fixed by a torsion constraint persists. An interesting idea to overcome it was presented in \cite{Butter:2021dtu}. It introduces an additional gauge symmetry which acts by a shift on all undetermined contributions in connections. Physical relevant quantities are then those that are invariant under this new, additional gauge symmetry. Closure of the new gauge transformations require to add more and more of them which triggers the transition from the double Lorentz group as gauge group $H$ to a Maxwell-$\infty$ algebra. This elegant solution should also have some interesting consequences for generalised Cartan geometry. Shift of the connections $\Omega^\beta_A$ and $\rho^{\alpha\beta}$ can be implemented by the additional constants $f_{\alpha B}{}^\gamma$ and $f_{\alpha}{}^{\beta\gamma}$ in \eqref{eq:shiftOmega} and \eqref{eq:shiftRho} respectively. As remarked in section~\ref{chap:GeneralisedCartanGeometryCurvature}, in this case the decomposition of the model algebra is not reductive. Already in Cartan geometry a reductive decomposition is typically assumed. Therefore, we did not gave further attention to this case here. However, it might be worth to be further analysed as it might be central for the elimination of undetermined components in generalised geometry.
 
Another possible avenue of research is concerned with higher dimensional ($>$2) dimensional $\sigma$-models whose Hamiltonian formulation is closely related to exceptional generalised geometry \cite{Hatsuda:2012vm,Hatsuda:2012uk,Hatsuda:2013dya,Duff:2015jka,Sakatani:2020iad,Osten:2021fil,Osten:2024mjt}. Latter extends the duality group from O$(d',d')$ to the exceptional groups E$_{d(d)}$ and thereby captures U-dualities as they appear in the context of M-theory. Recently, an adapted version of the double field theory construction which underlies this work has been presented for these larger duality groups \cite{Hassler:2023axp}. In particular, it shows that while for O$(d',d')$ only $\rho^{\alpha\beta}$ appears as additional connection besides the spin-connection $\omega^\beta_A$ in general there is a hierarchy of them organised by the so-called tensor hierarchy algebra \cite{deWit:2008ta,Riccioni:2009xr}. This phenomena is rooted in the existence of gauge symmetries for gauge symmetries, a central property of generalised geometry which it shares with higher Lie algebras. It would be interesting to define an exceptional version of our generalised Cartan geometry and study how it organises the phase space and Hamiltonian of higher dimensional sigma models. These models are much hard to access and thus widely unexplored. New mathematical tool will be crucial to access them.

Eventually, it is very tempting to leave the classical regime we have discussed here. Especially because the Hamiltonian formalism discussed here is one of the canonical routes to a quantum theory. The only determined components of the generalised Riemann tensor that contribute to the left-hand side of \eqref{eqn:masterCurrentAlgebra} is the generalised Ricci tensor. If it vanishes, one can integrate out the $H$-connection $A$ and is left with the Dirac bracket \eqref{eq:DiracBracket} as starting point for the canonical quantisation. Remarkably, a vanishing Ricci tensor also implies the vanishing of the one-loop $\beta$-functions and renders the model conformal, at least at this order in perturbation theory. For special examples \cite{Lacroix:2023ybi} has shown that the quantisation of \eqref{eq:DiracBracket} is indeed possible in this well-controlled setting and leads to vertex-algebra, like the W-algebra. An immediate question in the context of our article would be if generalised Cartan geometry continues to effectively capture the target space geometry and its quantum corrections or whether it needs to be modified. For another recent perspective on quantum corrections from double field theory for specific backgrounds see for example \cite{Lunin:2024vsx}.

\section*{Acknowledgements}

We thank Alex Swash for comments on the draft and Achilles Gitsis, Yuho Sakatani and Luca Scala for many enlightening discussions on the Poláček-Siegel construction. F.H. is supported by the SONATA BIS grant 2021/42/E/ST2/00304 from the National Science Centre (NCN), Poland. The work of O.\,H.\ was supported by the FWO-Vlaanderen through the project G006119N and by the Vrije Universiteit Brussel through the Strategic Research Program ``High-Energy Physics''. The research of D.O. is part of the project No. 2022/45/P/ST2/03995 co-funded by the National Science Centre and the European Union’s Horizon 2020 research and innovation programme under the Marie Sk\l odowska-Curie grant agreement no. 945339.

\vspace{10pt}
\includegraphics[width = 0.09 \textwidth]{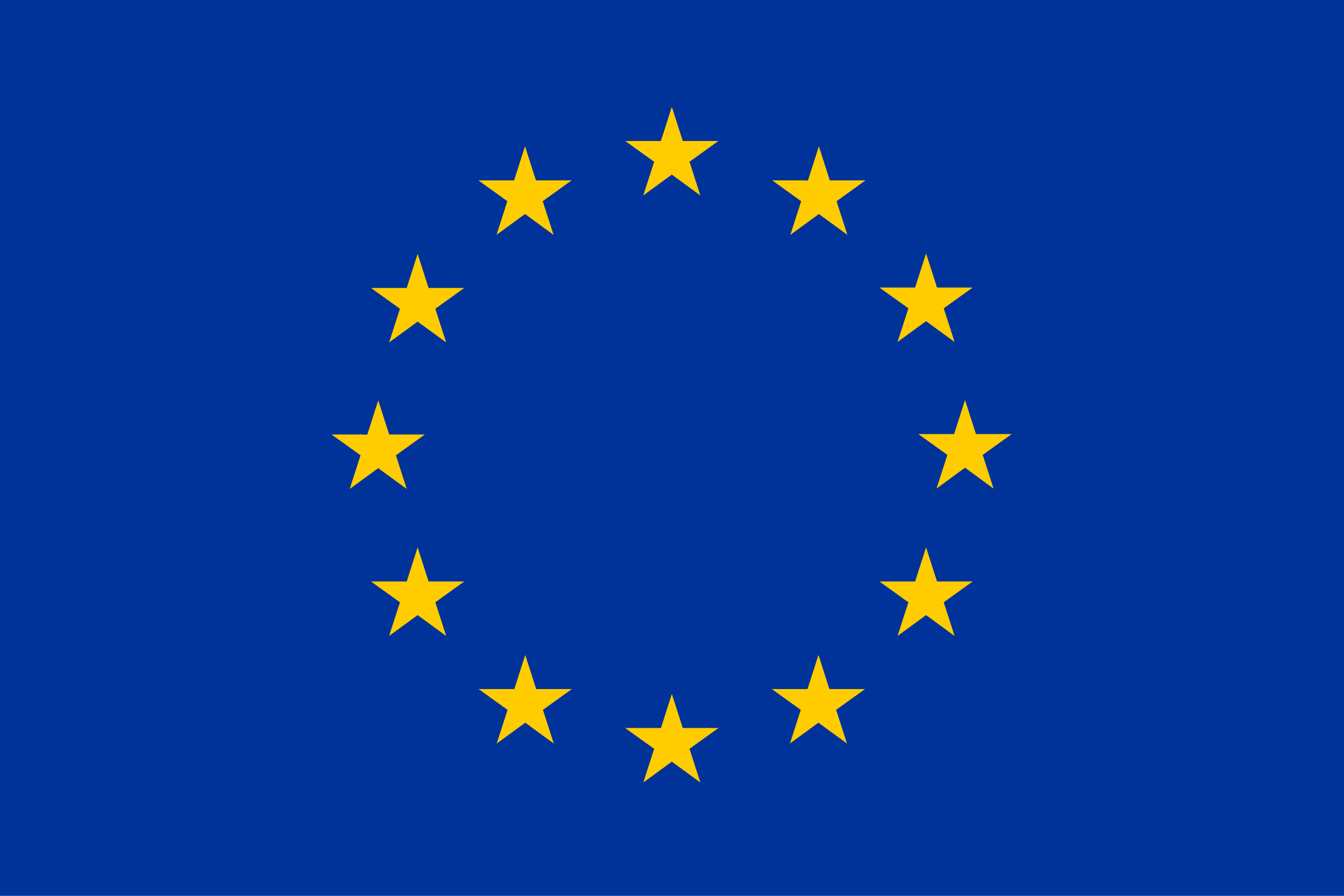} $\quad$
\includegraphics[width = 0.7 \textwidth]{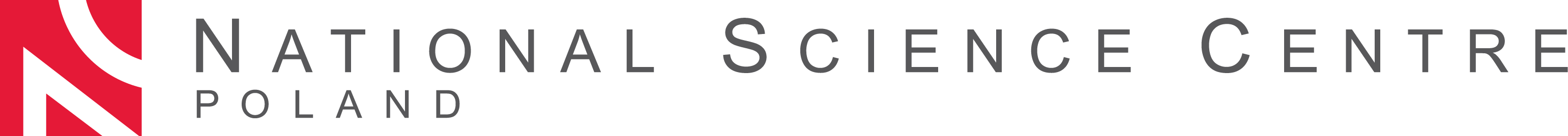}

\appendix

\section{Index dictionary}
\begin{itemize}
\item $\alpha,\beta,\gamma,...$ and $\kappa, \lambda, \mu ,...$: indices on $\mathfrak{h}$ resp. $TH$
\item $a,b,c,...$ and $k,l,m,...$: curved resp. flat indices on $TM$, transition with frame ${e_a}^m$
\item $A =({}_a,{}^a),B,C,...$ and $K,L,M,...$: curved resp. flat indices on $(T \oplus T^\star)M$, transition with generalised frame ${E_A}^M$
\item $\mathcal{A} = (\alpha , a), \mathcal{B},\mathcal{C},...$: indices on $TP$ or $\mathcal{A} = ({}_\alpha , A , {}^\alpha), \mathcal{B},\mathcal{C},...$: indices on $(T\oplus T^\star)P$ 
\item $\mathcal{K},\mathcal{L},\mathcal{M},...$: indices on model space algebra $\mathfrak{g}$ or $\mathfrak{d}$
\end{itemize}

\bibliographystyle{jhep}
\bibliography{References}

\end{document}